\documentclass[fleqn,10pt]{wlscirep}
\usepackage[utf8]{inputenc}
\usepackage[T1]{fontenc}
\usepackage{booktabs}
\usepackage{multirow}

\title{Personalizing Driver Safety Interfaces via Driver Cognitive Factors Inference}

\author[a,1,*]{Emily S. Sumner}
\author[a,1,*]{Jonathan DeCastro} 
\author[a,1]{Jean Costa}
\author[a,1]{Deepak E. Gopinath}
\author[a,1]{Everlyne Kimani}
\author[a]{Shabnam Hakimi}
\author[a]{Allison Morgan}
\author[a]{Andrew Best}
\author[a]{Hieu Nguyen}
\author[a]{Daniel J. Brooks}
\author[a]{Bassam ul Haq}
\author[a]{Andrew Patrikalakis}
\author[a]{Hiroshi Yasuda}
\author[a]{Kate Sieck}
\author[a]{Avinash Balachandran}
\author[a,2]{Tiffany Chen}
\author[a,2]{Guy Rosman}

\affil[a]{Toyota Research Institute}
\affil[1]{These authors contributed equally}
\affil[2]{These authors contributed equally}
\affil[*]{Correspondences can be sent to emily.sumner@tri.global and jonathan.decastro@tri.global}

\begin{abstract}
Recent advances in AI and intelligent vehicle technology hold promise to revolutionize mobility and transportation, in the form of advanced driving assistance (ADAS) interfaces.  Although it is widely recognized that certain cognitive factors, such as impulsivity and inhibitory control, are related to risky driving behavior, play a significant role in on-road risk-taking, existing systems fail to leverage such factors.  Varying levels of these cognitive factors could influence the effectiveness and acceptance of driver safety interfaces. 

We demonstrate an approach for personalizing driver interaction via driver safety interfaces that are triggered based on a learned recurrent neural network. The network is trained from a population of human drivers to infer impulsivity and inhibitory control from recent driving behavior. Using a high-fidelity vehicle motion simulator, we demonstrate the ability to deduce these factors from driver behavior.  We then use these inferred factors to make instantaneous determinations on whether or not to engage a driver safety interface.  This interface aims to decrease a driver's speed during yellow lights and reduce their inclination to run through them.
\end{abstract}
\begin{document}

\flushbottom
\maketitle

Advancements in driver safety systems have the potential to save lives~\cite{Singh2015-vm, Bareiss2019-sp}. However, the effectiveness of these systems varies considerably among different drivers. These safety systems could benefit from targeting of cause of individual driver's dangerous driving behaviour, which is known to be affected by many different factors, including cognitive, social, and situational \cite{nhtsa2019}.  Among the cognitive factors that influence risky driving behavior are \textit{impulsivity}, which is the tendency to act without thinking, and \textit{inhibitory control}, which is the ability to suppress goal-irrelevant stimuli and behavioral responses~\cite{Munakata2011-ux}. Risky driving has been associated with higher self-reported impulsivity~\cite{Constantinou2011-qz,Dahlen2005-ts, Walshe2017-wq, Hayashi2018-ul,National_Research_Council2007-gb}, and with poorer inhibitory control in relevant laboratory tasks~\cite{Hatfield2017-bf,Jongen2011-wj,  Walshe2017-wq, Hayashi2018-ul,National_Research_Council2007-gb}. These cognitive factors also influence individuals' reactions to different types of interfaces~\cite{McDonald2018-ge, Montgomery2014-gh}. 

The significance of impulsivity and inhibitory control as a risk factors for vehicle accidents has led to the idea of new intervention strategies. Paaver et al.~\cite{paaver2013preventing} showed that even a brief impulsivity intervention could help to prevent speeding. However, while there are numerous driver safety interfaces available, there is a gap in the research regarding the influence of impulsivity and inhibitory control on drivers' responses to these interfaces. More specifically, studies have not adequately explored how to tailor the deployment of these safety interfaces to individual drivers, taking into account their unique levels of impulsivity and inhibitory control. This personalization is crucial, as it can determine the effectiveness of the interface in enhancing driver safety.  

Thus, the efficacy of driver safety systems may vary in regards to individual differences in cognition. The design of human-machine interfaces, with a focus on incorporating specific cognitive characteristics, has the potential to enhance both their safety effectiveness and user acceptance~\cite{horberry2018driver}. Furthermore, the ability to estimate cognitive characteristics from observed driver behavior lays the groundwork for more personalized and effective safety interventions. 

We hypothesize that it is possible to construct effective driver assistance systems by learning neural representations that capture the cognitive factors of individual drivers. Building such a system that is geared towards inferring cognitive factors would allow us to fully separate \emph{how} we personalize the driver interface from \emph{what} we can personalize about the interface. In this way, we can build assistance systems that are highly versatile - for instance, if a new interface is developed, these can be integrated without additional re-training of the representation.  Such neural representations for cognitive factors will enable refinement of the estimated factors, as well as deployment of personalized safety intervention, at a large scale.

In this paper, we present experimental evidence of how factors such as impulsivity and inhibitory control can influence people's responses to driver safety interfaces and how the inference of such cognitive measures enable an approach for personalizing safety interfaces. We do so by constructing a neural network model that embeds driver behavior into a latent space that captures these factors, and demonstrate that embedded representation's utility for deciding on assistive driving interfaces deployment. The interfaces that we built are targeted towards inhibitory control and impulsivity. To our knowledge, we are the first to demonstrate driver assistance personalization in a high-fidelity simulator. In this paper we contribute 1)  A neural network model that capable of embedding individual cognitive factor differences based on recent driving behavior.  2) A decision-making system capable of personalizing assistive driver interface choices based on the inferred cognitive factors. 3) Experimental evidence of how impulsivity and inhibitory control relate to performance under different driver assistance choices on a new dataset collected in a large-scale, high-fidelity, driving simulator. 4) Demonstration in that dataset of how the neural network model affords conditioning of driver behavior via a personalized interface choice.

\section*{Related Works}
Our work is at the intersection of two active research areas: the role of cognitive factors in understanding driving behavior, and learning approaches 
for human-machine interaction.

\textbf{Cognitive Factors and Driving Behaviors}
Common approaches for assessing driving behavior commonly involve self-report surveys~\cite{af2011manchester}, ticketed speeding violations\cite{OBrien2013-cl}, or crash records\cite{Chang2014-ut}. While these measurements can be good indicators of risky driving behavior, self-report metrics such as these are not always reliable~\cite{Gemming2014-on}, contain private information, and do not lend themselves to seamless integration into preventative use with drivers. Other studies have shown driving characteristics can be estimated by measuring reactions to predetermined unsafe events in a simulated driving task~\cite{Hatfield2017-bf}. Our work uniquely examines a general approach to infer latent cognitive factors from behavior logs via a neural network encoder, and uses a high-fidelity driving motion simulator where behavior is closer to real vehicles than in lower-fidelity simulators (e.g., bench set up with a steering wheel)~\cite{schrum2023maveric}. 

\begin{figure}
    \centering
        \includegraphics[width=0.7\linewidth]{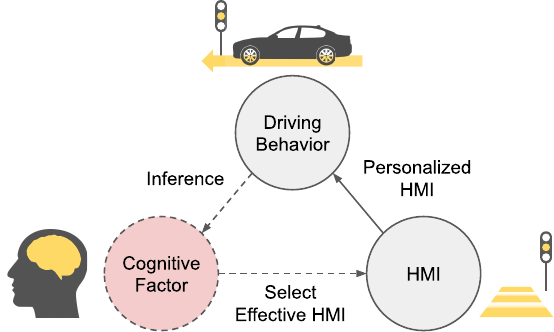}
    \caption{A conceptual overview of our framework. Latent factors embed cognitive measures from the driving behavior, and used to inform HMI choice(dashed lines). Solid line marked the observable driving behavior and personalized HMI.}
\label{fig:flowchart}
\end{figure}

In addition to measuring driving behavior, researchers often measure impulsivity and other behavioral and cognitive factors via tests and questionnaires~\cite{Diamond2013-pe}. However, for these cognitive factors to effectively enhance vehicle safety systems, they should be estimated in a scalable way and applied to the development of personalized assistive interfaces within vehicles. Estimating these factors from driving behavior could lead more accurate information about drivers and further lead to effective intervention design and deployment criteria.

\textbf{Learning Latent Factors for Human-Machine Interaction}
An intelligent vehicle being a robotic system, our approach relates to efforts in personalizing interactions between humans and robots or other machines.
Prior work in machine learning for human-machine interfaces and human-robot teaming has focused on various human-robot interaction modalities such as driver monitoring, optimal shared control laws, and design of assistive robot behaviors (see e.g.~\cite{Kaplan2015-ad,Schaff2020-bm,Losey2022-qw,backman2022reinforcement}). However, these approaches for human-robot interactions typically do not explicitly consider individual differences in cognitive factors and therefore fall under the category of a ``one-size-fits-all'' design. 

The same is true for modern-day driver assistance systems such as lane-departure warnings or forward-collision warnings. Typical interventions issued by such systems depend on an individual's state and action history and manifest as corrections of unsafe or suboptimal human actions generated from a policy learned from a desired set of behaviors required of the system~\cite{Nidamanuri2022-qf}. Such approaches have been found to over-fit to the average-case behavior of individuals in a population, leading to incorrect inference of the human’s state and poor generalizability~\cite{Xie2020-aq, Tsividis2021-rf}. 
Recent work has shown that learning latent representations summarizing human behavior can improve teaming and interaction with the human. For instance, work on dialog systems~\cite{Yang2021-so}, recommender systems~\cite{Song2017-qt, Yu2019-vv}, and intent recognition for products and motion~\cite{Tanjim2020-mb, Rudenko2019-yq} have demonstrated that latent representations are capable of better predicting the user's need for a given intervention and their reaction to that intervention.  We posit that using this representation as a basis for deciding whether to interact and which modes of interaction to use should improve safety over ``one-size-fits-all'' decision schemes. 

In this paper, we explore how to personalize human-machine interfaces based on people’s impulsivity and inhibitory control. We posit that latent factors such as impulsivity and inhibitory control can be inferred in an automated manner from driving behavior and can inform choices of interactions with the drivers to benefit them at a large scale.

\section*{Contrastive Learning of Personal Traits}
\label{subsec:comp_modeling}
We now proceed to describe our approach for encoding latent cognitive factors. The resulting neural network distills a human driver's recent driving history down to a low-dimensional parameter space whose structure can be easily shaped via multiple cognitive measures in a semi-supervised manner. The model we use includes a context encoder whose input is a time-receding, fixed-window trajectory of driving behavior in a scenario and whose output is a low-dimensional latent vector. This latent representation is then coupled with a separate decision-making module that takes in this latent vector and outputs a decision at each decision time-step; for instance, whether or not to present a particular HMI to the driver at the current time-step. The architecture is shown in Fig.~\ref{fig:arch}, with further details in the supplemental information.  As a result of experimentation, we found that a two-dimensional latent vector provided sufficient capacity to capture relevant cognitive factors, yet allow direct interpretation of the learned trends in the representation without possible distortions introduced by dimensionality reduction schemes (e.g. t-distributed Stochastic Neighbor Embedding~\cite{van2008visualizing}).

The context encoder is a neural network, $q_{\psi}(z \mid \tau)$, defining the probability of latent vector $z$ given a past trajectory $\tau$ of the driver.  
We represent $q_{\psi}(z \mid \tau)$ as a long short-term memory (LSTM) network~\cite{Hochreiter1997-fv}, taking in a trajectory history.  The hidden layer $h$ is fed into two linear layers that output the mean and log-variance of the latent encoding~\cite{kingma2013auto}.  

As driving actions do not directly relate to psychological traits, we leverage contrastive learning ~\cite{Gutmann_undated-lh,Khosla2020-ks} to encourage the latent representation to conform to measured cognitive factors (we introduce the specific factors we use in the Results section). Because decisions should be based on more than one cognitive factor, we consider our \textit{cognitive factor target} to be a vector.

The context encoding model transforms a driver's past driving history $\tau$ to a latent vector $z$, and uses a decoder network $p_{\theta}(a|z)$ to predict the driver's action $a$ at the current time-step.  We set up the loss terms to encourage $z$ to capture both the individual's cognitive factors and reconstruction of driver actions, with the factors allowing for the downstream decision-making module to have awareness of any time-independent factors inherent to the individual driver, and driver actions allowing for awareness of the behaviors in a given situation.  Any scene context information present in $\tau$ will indirectly manifests in $z$ through $q_{\psi}(z \mid \tau)$.  Thus, we expect a weak dependence of predicted driver action on scene context.  The overall loss used to train the encoder consists of three components:
\begin{itemize}
    \item $L_1(a, z; \theta) = -\mathbb{E}_z \log p_{\theta}(a \vert z)$ 
    is the likelihood of action $a$ under the model (reconstruction loss) induced by the conditional distribution $p$ over $z$, where $z$ characterizes driving behavior up to time $t$ and $\theta$ represents the parameters of the action decoder network.
    \item $L_2(z,y)$, a contrastive loss supervised using a vector of cognitive factor targets $y$~\cite{rai2021cocon}.  For continuous-valued cognitive measures, this loss is 
    \begin{align}
    L_2(z,y; \psi) =& \sum_{z' \in \mathcal{Z}} (1 - \Vert y_{z} - y_{z'} \Vert^2) \ell(z, z')^2 \nonumber + \Vert y_{z} - y_{z'} \Vert^2 \max(0, \epsilon - \ell(z, z'))^2,
    \end{align} 
    where $\mathcal{Z}$ represents a training samples batch, where each independently-sampled $z, z' \in\mathcal{Z}$ is a $\vert Z \vert$-dimensional latent vector induced by the LSTM context encoder with parameters $\psi$, $y_{z}$ is a vector of batch-normalized cognitive measures associated with $z$, $\ell(z, z')$ is a measure associated with two vectors $z$ and $z'$ (which we choose as their Euclidean distance, i.e.~$\Vert z - z'\Vert$), and $\epsilon$ controls the magnitude of dissimilarity of $y$-values in $z$-space, where a larger $\epsilon$ enforces higher separation of $\Vert z - z'\Vert$ for fixed $\Vert y_{z} - y_{z'}\Vert$. 
    \item $L_3(z) = D_{KL}(q_{\psi}(z \mid \tau)\vert \mathcal{N}(0,I))$, a Kullback–Leibler (KL)-regularization loss for the distribution of $z$, e.g. as in~\cite{Kingma2013-mr,Rezende2014-mq}. $\mathcal{N}(0, I)$ is the unit-normal distribution of appropriate dimension.
\end{itemize} 
These terms are combined into an overall training loss:
\begin{align}
    L(a, z, y; \theta) = \alpha_1 L_1(a, z; \theta) + \alpha_2 L_2(z, y; \psi)+ \alpha_3 L_3(z)
\end{align}
where $\alpha_1$, $\alpha_2$, and $\alpha_3$ are the respective loss coefficients.
\begin{figure}
    \centering
        \includegraphics[width=0.95\linewidth]{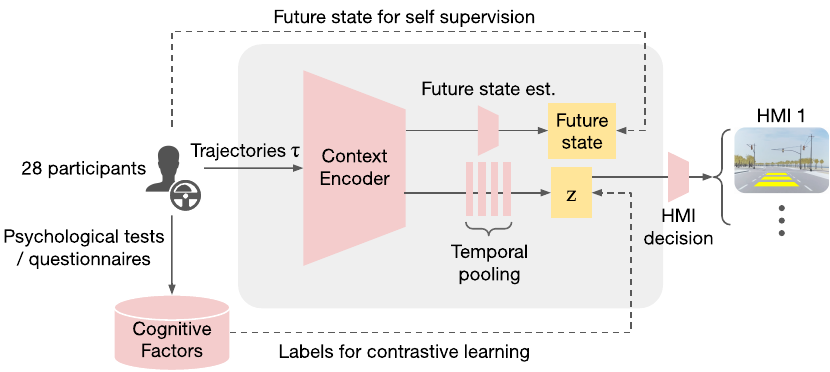}
    \caption{Overall system architecture, including context encoder, decoder for future state and action prediction, outputs of cognitive measures, and latent factors used for human-machine interface selection and decision-making.}
\label{fig:arch}
\end{figure}

We evaluate the utility of the inferred latent factors model by marrying it with an decision rule for selecting the activation of the human-machine interface (HMI). The decisions are defined via a simple classifier whose inputs are the inferred latent factors. The classifier is set to optimize a criteria for HMI selection within the training data. We take the criteria for classification to be the difference in average speed when the yellow light is active, both with and without HMI (averaged across trajectories for a single subject).  This criterion reflects the speed reduction induced in the subject when an HMI is provided to the driver.  We use Support Vector Regression~\cite{Chang2011-xe} with a polynomial kernel as our decision model.

\section*{Experiment}
\label{sec:experiment}
\begin{figure}
     \centering
         \includegraphics[trim={0 2cm 0 2cm}, clip, width=0.7\linewidth]{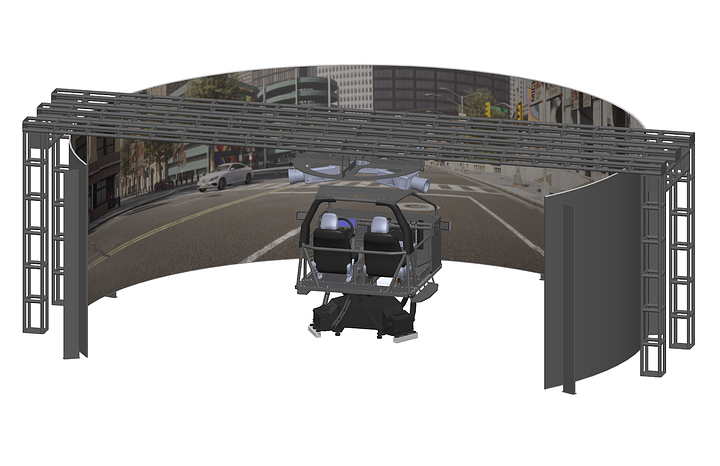}
     \caption{An illustration of the driving motion simulator used for data collection.}
 \label{fig:motionsim}
 \end{figure}

To test our approach, drivers were recruited and asked to execute multiple driving trials (laps) on a closed-loop road section in a high-fidelity full-scale vehicle motion simulator, described in details in~\cite{schrum2023maveric}.  An illustration of the driving motion simulator is shown in Fig.~\ref{fig:motionsim}.
Participants drove on a road with traffic lights that randomly changed from green to yellow at varying times of arrival of the vehicle at the traffic light, inducing a zone of dilemma~\cite{Zhang2014-qq}. 
We collected four driving trials (laps) where participants interacted with different prototype driver safety interfaces and one baseline driving trial, without the interfaces. Data from the baseline trials that was used for latent factor inference via a learned neural network model. 

\subsubsection*{Driver Safety Interfaces}
\begin{figure}[tbp]
    \centering
    \begin{minipage}{.24\textwidth}
        \centering
        \includegraphics[width=1\linewidth]{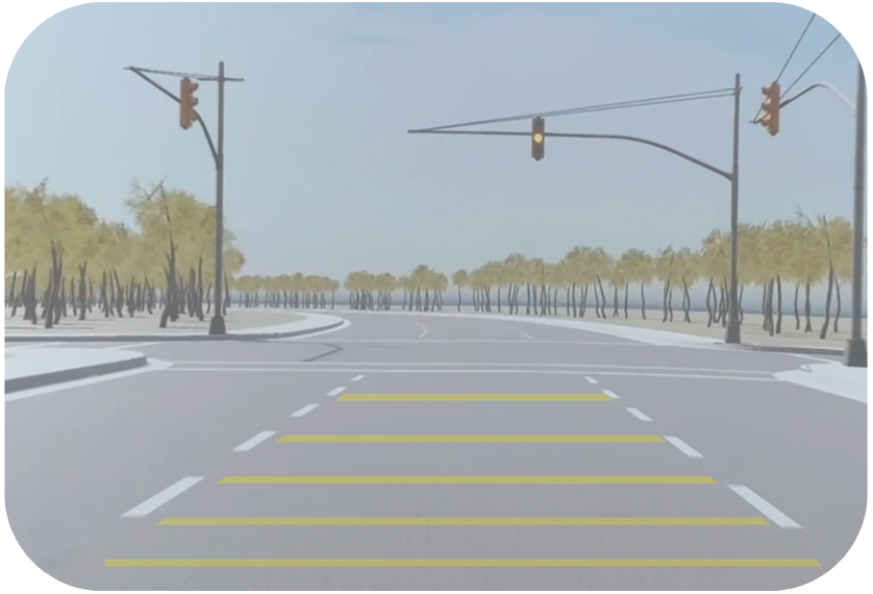}
        
        a
    \end{minipage}
    \begin{minipage}{0.24\textwidth}
        \centering
        \includegraphics[width=1\linewidth]{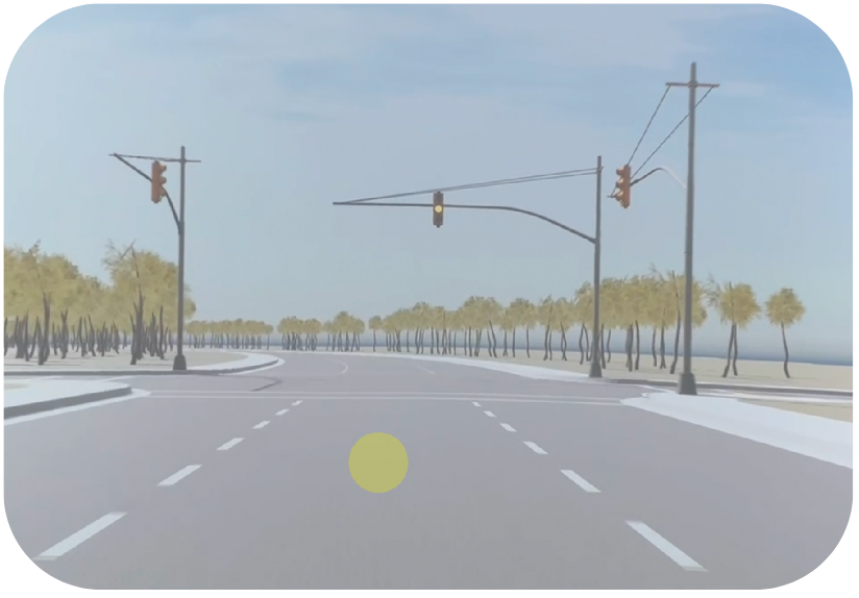}
        
        b
    \end{minipage}
    \begin{minipage}{0.32\textwidth}
        \centering
        \includegraphics[width=1\linewidth]{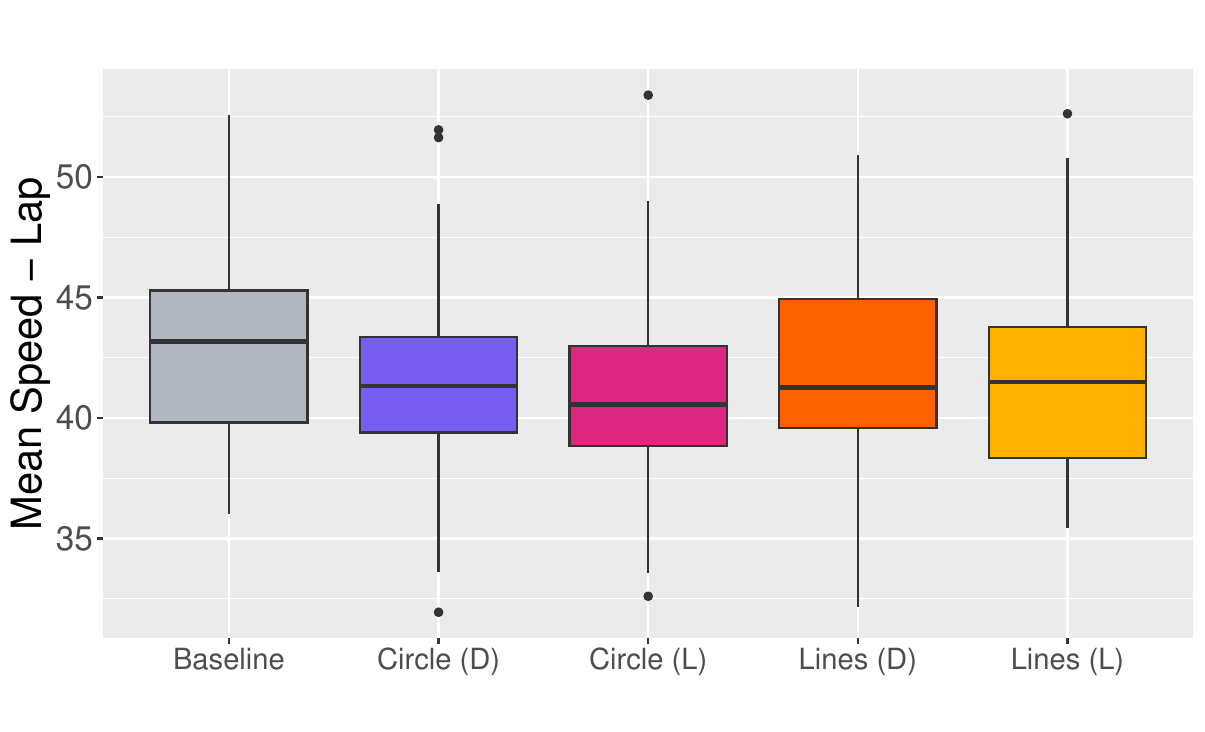}
        
        c
    \end{minipage}
    \caption{Example of HMI types used in the experiment when nearing a traffic light. a) transverse markings. b) yellow circle. c) the effect of different choices of HMI types on the mean speed at yellow traffic lights. }
\label{fig:HMI_images}
\end{figure}

Two types of warning interfaces were used:  a) transverse markings, projected on the road the car was driving; and b) a 2D yellow circle, projected as if it appeared in a heads-up display. Fig.~\ref{fig:HMI_images} shows the virtual scenario and both interface types. For each interface, we also manipulated the trigger condition to display it. Each interface was either displayed when the vehicle approached the traffic light (185 meters away) or when the upcoming traffic light changed from green to yellow.

\subsubsection*{Participants}
Thirty nine Northern California-based drivers 18 and older (\textit{Mean age = 49, Female = 16, Non\_binary = 1}) were recruited to participate in our study via Fieldwork, a global market research firm. Participants were excluded from participation if they had an inactive driver’s license, were pregnant or didn't have proof of COVID-19 vaccination. (See details in the recruitment section in the supplemental information.)  This research was reviewed, approved, and done according to the human-subject guidelines set by the Western Institutional Review Board-Copernicus Group (WCG) IRB. Participants filled out a consent form prior to participation and were compensated \$150 for their two hour participation. 

Half of the participants were between the ages of 18-22, the other half were over the age of 65. We recruited from these two populations since we believed they would lead to the most diverse sample of driving data~\cite{National_Research_Council2007-gb, jonah1990age}. While age-related differences are not discussed in this paper, additional analyses can be found in the supplemental. 

Of these 39 participants, 7 participants did not complete the driving trials due to motion sickness. Of the 32 remaining participants, the data of 5 participants was excluded from the analysis due to technical or operational issues.

\subsubsection*{Measures}
\begin{itemize}
    \item \textbf{Impulsivity: }
   
    To assess participants' impulsivity~\cite{carver1994behavioral}, we used the \textbf{BIS/BAS} scale and the \textbf{UPPS-P} scale. The BIS/BAS was used to measure both the behavioral inhibition system (BIS) and the behavioral activation system (BAS), while the UPPS-P was used to account for the different facets of impulsivity. 
    \item \textbf{Inhibitory Control:}
    We used the \textbf{Go-No Go task}~\cite{gomez2007model} and the \textbf{Stop Signal task }\cite{lappin1966use,verbruggen2019consensus} to measure response inhibition. To obtain the measures from the Stop Signal task, we followed the recommendations from Verbruggen et al.~\cite{verbruggen2019consensus}.
    
    \item \textbf{Self-reported Driving Behavior: } To assess participants' road errors and violations, we used the Manchester Driver Behavior Questionnaire (DBQ)~\cite{af2011manchester}.   
    \item \textbf{Driving Behavior in the Motion Simulator: } We also captured driving behavior as participants drove in the motion simulator. We recorded their driving speed, acceleration and response to yellow traffic lights.
\end{itemize}
We only consider this partial set of cognitive measures as the supervision signal to shape  our model's latent space, as a full set of measures will be impractical in the fleet-level case we envision.

\section*{Results}
We analyzed the relationship between various aspects of impulsivity, inhibitory control, driving behavior, and responses to human-machine interfaces designed to encourage drivers to slow down. We then analyzed the performance of our model in inferring participants' cognitive factors and in predicting if they should interact with our human-machine interfaces.

\subsection*{Relationship Between Cognitive Factors and Driving Behavior}

To understand the relationship between the different cognitive factors and driving behavior when reacting to the yellow lights, we conducted a Bayesian correlation analysis using the JASP software~\cite{JASP2024}. For the analysis, we used the data of all driving laps -- including the ones with HMIs presented. A table with all of the Bayesian correlations can be found in the Supplementary Information document. As shown in these tables, a number of significant correlations emerged. 

The self-reported ordinary violations measure on the DBQ was positively correlated with the mean speed at the yellow light (r=0.4, BF10=9693) and the maximum speed when the yellow was active light (r=0.54 BF10=$1.141\times 10^9$), indicating that drivers who reported higher levels of ordinary violations were more likely to speed through yellow lights in this task. 

We found several correlations between the BIS/BAS measures and driving behavior. In particular, BAS Fun Seeking was positively correlated with the mean speed at the active yellow light (r=0.473, BF10=$1.700\times 10^6$) and the maximum speed at the yellow light (r=0.31, BF10=99.19). This suggests that individuals who have a higher desire for new and exciting experiences may be more likely to take risks while driving, such as speeding through yellow lights. BAS Reward Responsiveness was also positively correlated with the maximum speed at an active yellow light (r=0.29, BF10=39.63). 

Similar to the BIS/BAS measures, various correlations emerged using the UPPS-P subscales. For instance, UPPS-P Positive Urgency was positively correlated with the maximum speed at an active yellow light (r=0.28, BF10=26.93), and UPPS-P Sensation Seeking was positively correlated with the mean speed at the active yellow light (r=0.29, BF10=42.89) and the maximum speed at the active yellow light (r=0.47, BF10=$1.540\times 10^6$). These results are consistent with the results found for BAS Fun Seeking and BAS Reward Responsiveness, which provides further evidence that people who desire fun, new and thrilling experiences are more likely to speed and take risks when reacting to traffic lights.

Multiple correlations also emerged using the measures from the Stop Signal task. For instance, the reaction time on go trials with a response (goRT\_all) was negatively correlated with the mean speed at the yellow light (r=-0.38, BF10=2933). This suggests that drivers with longer reaction times may be more likely to slow down at yellow lights rather than speeding through them.

Finally, we also found numerous correlations using the Go/No-Go measures. Among the correlations, the average response time (gonogo\_average\_rt) was negatively correlated with the mean speed at the yellow light (r=-0.46, BF10=352747) and the maximum speed at the yellow light (r=-0.40, BF10=9205), which is consistent with the reaction time results from the Stop Signal task (e.g. goRT\_all).

\begin{figure}[tbp]
    \centering
    \begin{minipage}{\textwidth}
        \centering
        \includegraphics[trim={0 6.2cm 0 0}, clip, width=0.5\linewidth]{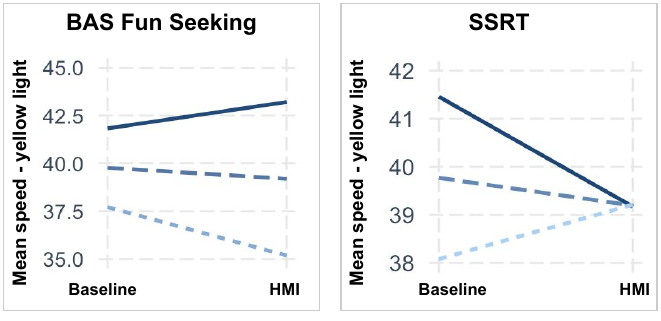}~
        \includegraphics[trim={0 0.5cm 0 5.5cm}, clip, width=0.49\linewidth]{interaction_plots_final_.pdf}
        \includegraphics[trim={0 0 0 11cm}, clip, width=0.6\linewidth]{interaction_plots_final_.pdf}
    \end{minipage}
    \caption{Interaction plots showing how the presence of the HMI interacted with different factors. From top left to bottom right: a) BAS Fun Seeking: Motivation to find novel rewards spontaneously; b) SSRT: Stop Signal Reaction Time; c) UPPS-P - Positive Urgency: Tendency to act impulsively due to positive affect; d) DBQ Ordinary Violations: Self-reported ordinary driving violations.}
\label{fig:interaction_plots}
\end{figure}

\subsection*{Impact of Cognitive Factors on People's Responses to the Interfaces}

We fitted separate linear mixed models to predict each driving behavior measure based on interface condition (Table~\ref{tab:model_effects}. All conditions demonstrated a statistically significant and negative effect on the mean speed during the lap, as depicted in Fig.~\ref{fig:HMI_images}(c).

To further understand how different factors affect drivers' responses to HMI, we conducted a linear mixed models (LMM) analysis, using multiple LMMs to examine the effects of various factors, including the presence or absence of HMI (\textit{HMI\_presence}) and their potential interactions. Participant ID was used as a random effect to account for individual differences. The \textit{lmer} function in the \textit{lme4} R package~\cite{lme4} was employed for predicting mean speed when yellow lights were active based on these variables as

$\textrm{Mean\_speed\_yellow} \sim \textrm{HMI\_presence}*\textrm{Cognitive\_Factor} + (1 | \textrm{Participant}),$ 

\noindent where $(1 | \textrm{Participant})$ denotes the random intercept. The models were fitted using the Restricted Maximum Likelihood (REML) estimation method, and the t-tests utilized Satterthwaite's approximation method.

For detailed statistical outcomes of all measures, please refer to Table 1. For a visual representation of some interaction effects, please see Fig.~\ref{fig:interaction_plots}, which complements the textual analysis. Here, we highlight some key findings that were noted to have a strong effect:

\vspace{\baselineskip}

\textbf{BIS/BAS}:
The BAS Fun Seeking scale showed a significant main effect of HMI presence ($\beta=-11.14, SE=4.64, t=-2.4, p=0.01$), indicating that individuals with higher scores were likely to drive faster at yellow lights when HMI was present. Additionally, a significant interaction between HMI presence and BAS Fun Seeking was observed ($\beta=0.9, SE=0.39, t=2.31, p=0.02$).

\textbf{UPPS-P}:
The Positive Urgency subscale revealed a significant main effect of HMI presence ($\beta=-7.6, SE=3.06, t=-2.48, p=0.01$), along with an important interaction with HMI presence ($\beta=1.23, SE=0.38, t=3.22, p=0.002$), suggesting that those with higher levels of positive urgency increased their speed at yellow lights in the HMI presence condition.

\textbf{Go/No-Go Measures}:
The analysis revealed a significant main effect of average response time (gonogo\_average\_rt) on mean speed, with a negative influence ($\beta=-0.07, SE=0.027, t=-2.56, p=0.01$). No significant interaction effects were observed in these measures.

\textbf{Stop Signal Measures}:
Significant main effects were observed for goRT\_all ($\beta=-0.02, SE=0.008, t=-2.68, p=0.01$) and usRT ($\beta=-0.02, SE=0.01, t=-2.36, p=0.02$), along with a notable interaction between HMI presence and Stop Signal Reaction Time (SSRT) ($\beta=-0.01, SE=0.007, t=-2.00, p=0.04$). This suggests that individuals with higher SSRTs tended to drive slower after the yellow-light transition after they are exposed to the HMI.

\textbf{Manchester DBQ}:
For the Ordinary Violations subscale, a significant main effect of HMI presence ($\beta=-6.99, SE=2.75, t=-2.53, p=0.01$) and a significant interaction with HMI presence were found ($\beta=0.47, SE=0.19, t=2.44, p=0.01$). This indicates that those reporting higher ordinary driving violations were likely to drive faster after the yellow-light transition when exposed to the HMI. 

\begin{table}[tbp]
\centering
\caption{Summary of Linear Mixed Model Effects. The top part of the table details the main effects of various measures on the dependent variable, while the lower part shows the interaction effects between the Human-Machine Interface (HMI) presence and each measure.}
\label{tab:model_effects}
\begin{tabular}{lcccc}
\toprule
\textbf{Measure} & $\mathbf{\beta}$ & \textbf{SE} & \textbf{t} & \textbf{p} \\ 
\midrule
\multicolumn{5}{c}{\textbf{Main Effects}} \\
\midrule
BIS/BAS BAS\_reward & -0.03 & 14.39 & -3.15 & 0.002 \\
BIS/BAS BAS\_Fun & -11.14 & 4.64 & -2.4 & 0.01 \\
UPPS-P Positive Urgency & -8.71 & 2.65 & -3.28 & 0.001 \\
UPPS-P Sensation Seeking & -7.6 & 3.06 & -2.48 & 0.01 \\
Go/No-Go Measures gonogo\_average\_rt & -0.07 & 0.027 & -2.56 & 0.01 \\
Stop Signal Measures goRT\_all & -0.02 & 0.008 & -2.68 & 0.01 \\
Stop Signal Measures usRT & -0.02 & 0.01 & -2.36 & 0.02 \\
Manchester DBQ Ordinary Violations & -6.99 & 2.75 & -2.53 & 0.01 \\
Manchester DBQ Errors & -12.71 & 6.03 & -2.1 & 0.03 \\
\midrule
\multicolumn{5}{c}{\textbf{Interaction Effects}} \\
\midrule
BIS/BAS BAS\_reward * HMI\_presence & 1.48 & 0.47 & 3.095 & 0.002 \\
BIS/BAS BAS\_Fun * HMI\_presence & 0.9 & 0.39 & 2.31 & 0.02 \\
UPPS-P Positive Urgency * HMI\_presence & 1.23 & 0.38 & 3.22 & 0.002 \\
UPPS-P Sensation Seeking * HMI\_presence & 0.64 & 0.26 & 2.38 & 0.01 \\
Stop Signal Measures SSRT * HMI\_presence & -0.01 & 0.007 & -2.00 & 0.04 \\
Manchester DBQ Ordinary Violations * HMI\_presence & 0.47 & 0.19 & 2.44 & 0.01 \\
Manchester DBQ Errors * HMI\_presence & 0.95 & 0.46 & 2.03 & 0.04 \\
\bottomrule
\end{tabular}
\end{table}

\subsection*{Inferring Inhibitory Control and HMI Choice from Driving Behavior}

Given the various measures collected in the study, we used stepwise regression to select the most important features for training our model. We combined forward selection, starting with an empty model and adding the predictor that produced the largest increase in model fit, with backward elimination, removing the predictor that produced the smallest decrease in model fit until no further improvement was observed. By following this process, the stepwise regression yielded a set of four cognitive factors to be used in the model:  UPPS-P - Positive Urgency, BAS Fun Seeking, goRT\_all, and DBQ - Ordinary violations. 

\begin{figure*}
    \centering
    \begin{minipage}{0.6\textwidth}
        \centering
        \includegraphics[width=0.55\linewidth, trim=30 0 30 20, clip]{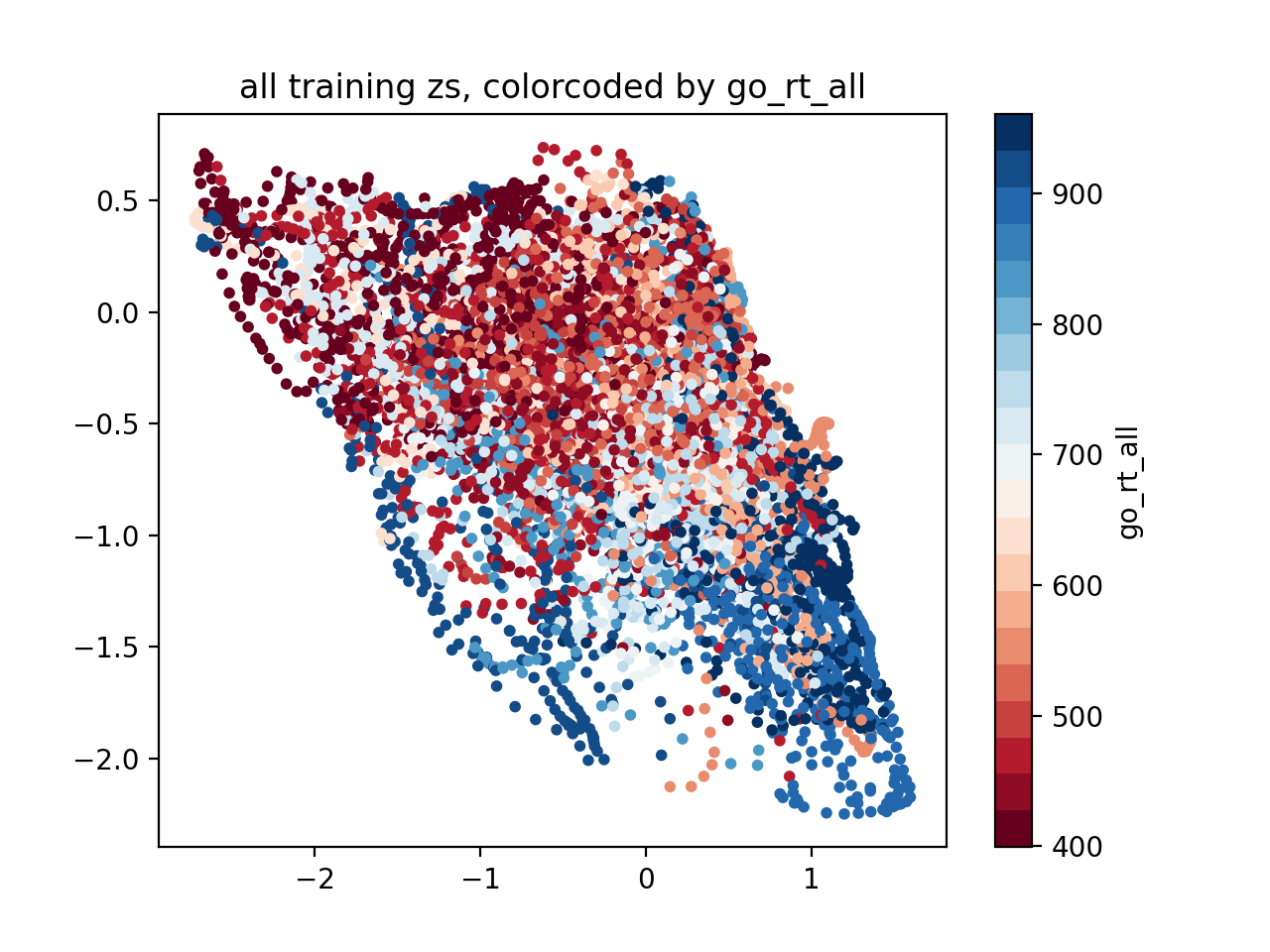}~
        \includegraphics[width=0.55\linewidth, trim=30 0 30 20, clip]{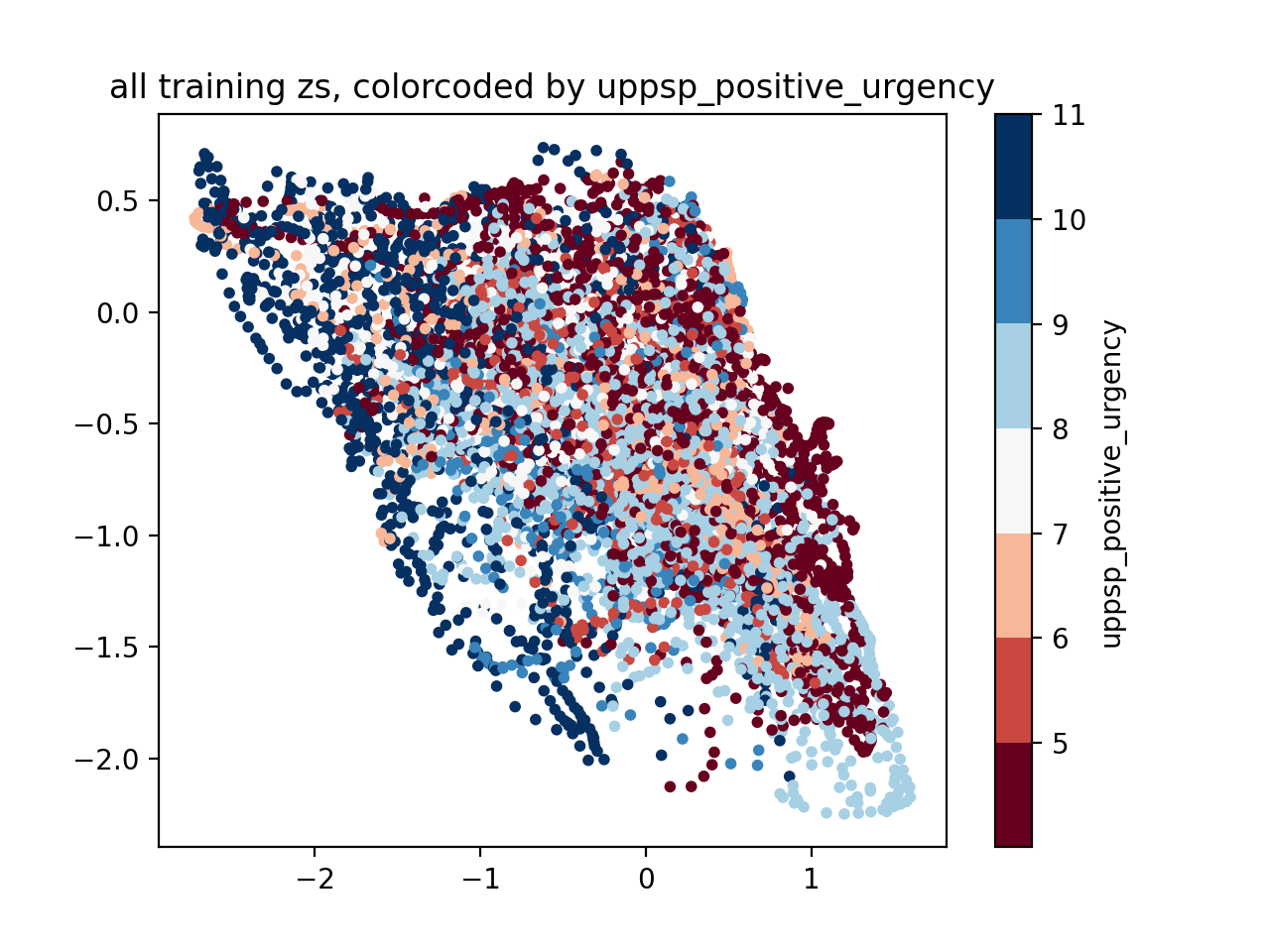} \\
        \includegraphics[width=0.55\linewidth, trim=30 0 30 20, clip]{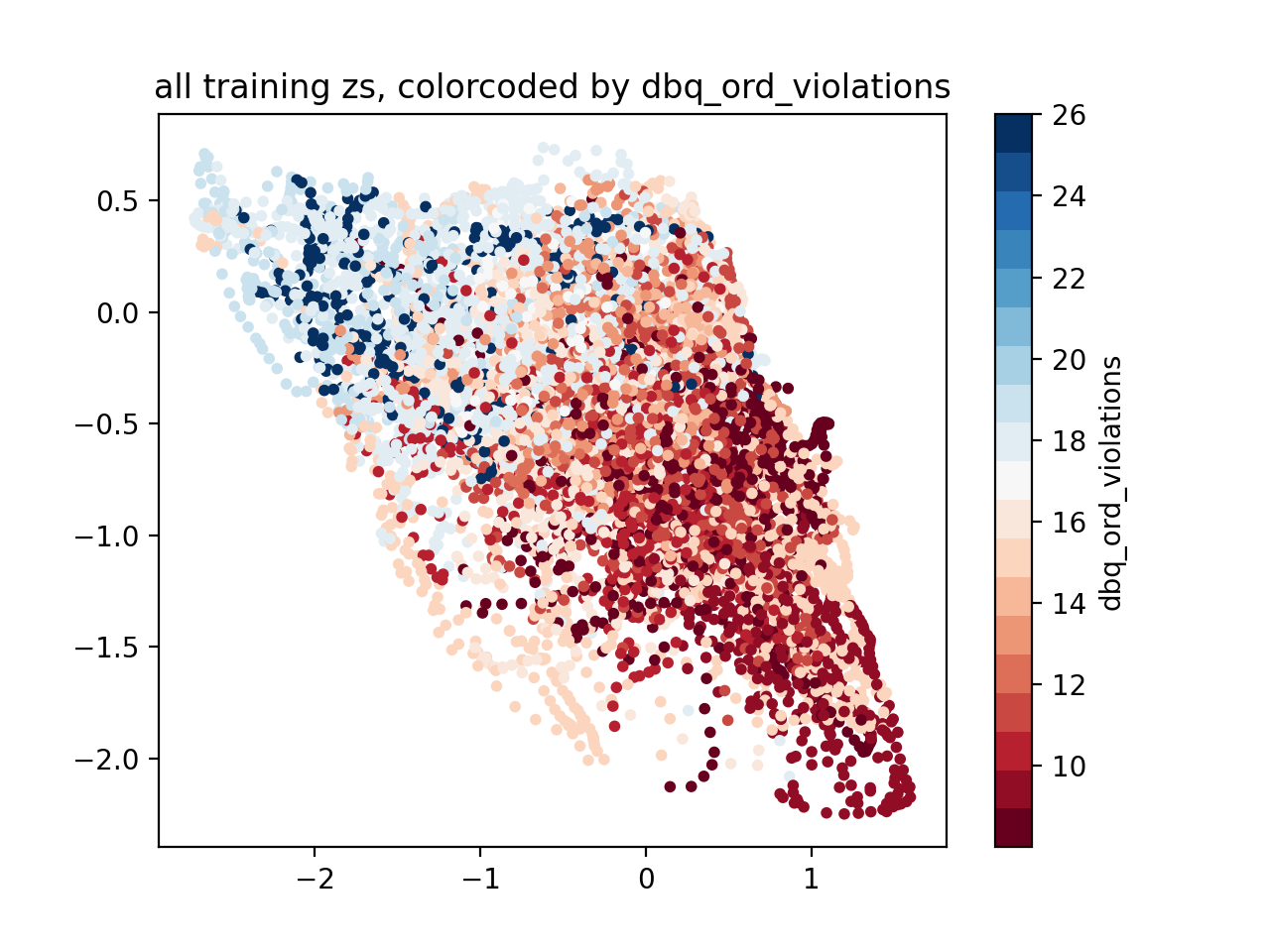}~
        \includegraphics[width=0.55\linewidth, trim=30 0 30 20, clip]{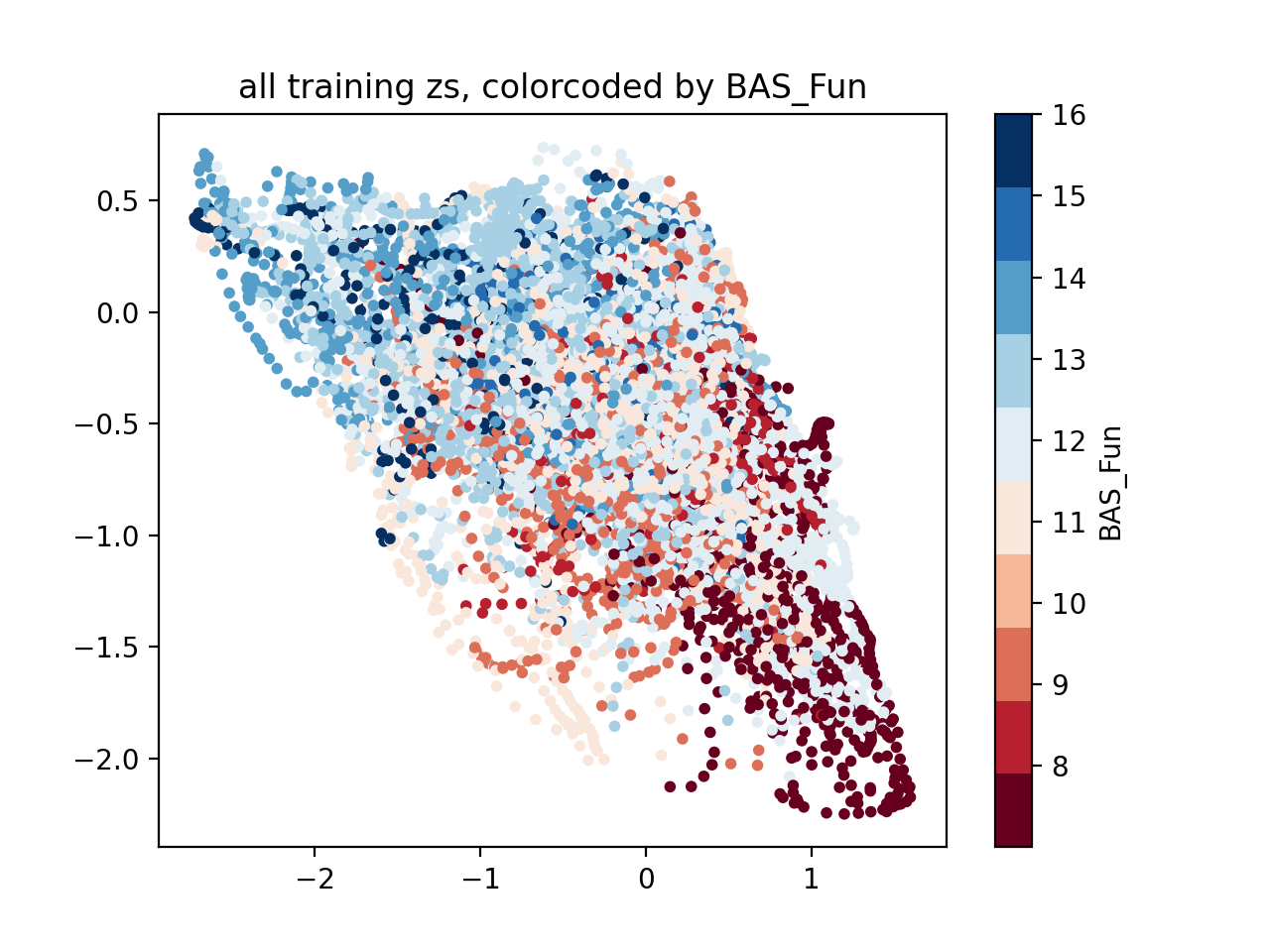} \\
        
    \end{minipage}%
        
    \caption{Example embedding and decision module result based on training data from a 27-subject fold. 
    a) Embedding of participants' past history trajectories with contrastive loss based on four factors: goRT all, UPPS-P Positive Urgency, DBQ Ordinary Violations, and BAS Fun. Colors mark low (red) to high (blue) measures. 
    }
\label{fig:latent_factors}
\end{figure*}

We adopt the learning approach described to infer cognitive factors based on the subjects' driving during the experiment. To evaluate the model on the available data, we use leave-one-out cross-validation over the 27 subjects, averaging model performance over 10 random seeds, and capture properties of the embedding and the resulting training decision criteria performance. We include a complete description of the training and evaluation steps and further findings in the supplemental information. The distribution of the inferred latent factors is shown in Fig.~\ref{fig:latent_factors}a. Qualitatively, we observe that fairly strong clustering has emerged for each of the cognitive factors which indicates the effectiveness of the the contrastive learning approach is effective. To quantify this further, we show in Table~\ref{table:kl_cfs} the fit between the distribution of the selected cognitive and the inferred latent factors. Since there is no direct or linear mapping assumed in contrastive learning, we probed the uniformity of the inferred embedding. We used the KL distance between the cognitive measures and the inferred factors' distribution. The results demonstrate the model's ability to infer several variables interest centered around impulsivity and inhibitory control.

\begin{table}[]
\centering
\caption{Normalized KL Divergences of the subjects for the cognitive measures used in the contrastive loss, averaged over 10 folds (higher is better).  We normalize over an ideal clustering result with two Normal distributions separated by a unit-distance (the regularization term $L_3$). Note BAS Fun is significantly higher, indicating stronger separation.}
\begin{tabular}{cccc}
\toprule
goRT all & \begin{tabular}[c]{@{}c@{}} UPPS-P Positive \\ Urgency \end{tabular} & \begin{tabular}[c]{@{}c@{}} DBQ Ordinary \\ Violations \end{tabular} & BAS Fun\\
\midrule
0.322 & 0.299 & 0.288 & 0.526  \\
\bottomrule
\end{tabular}
\label{table:kl_cfs}
\end{table}

\begin{table}[]
\centering
\caption{Resulting accuracy of interface selection based on the inferred latent factors using test datasets obtained by performing leave-one-out cross-validation on the full set of tests subjects. As can be seen, the inferred latent factors enable personalized HMI selection with 55\% balanced accuracy and 0.145 Cohen Kappa, compared to a balanced-random HMI choice 50\% and 0.001 Cohen Kappa, with the personalized scheme reducing yellow light driving speed by 0.59 m/s (standard error=1.58) compared with random.}

\begin{tabular}{lcccc}
\toprule
\multirow{2}{*}
{Decision Rule} & \multicolumn{2}{c}{\begin{tabular}[c]{@{}c@{}}Mean Yellow-Light \\ Speed (m/s)\end{tabular}} & \multirow{2}{*}{\begin{tabular}[c]{@{}c@{}}Cohen's \\ Kappa \\ Score\end{tabular}} & \multirow{2}{*}{\begin{tabular}[c]{@{}c@{}}Balanced \\ Accuracy\end{tabular}} \\
                               &                          & Standard               &                           &                                    \\
                               & $\mu$                    & Error                  &                           &                                    \\
\midrule
No-HMI                         & 17.36                    & 1.12   & 0.0           & 0.50                                \\
Always-HMI                     & 15.48                    & 1.10   & 0.0           & 0.50                                \\
Random                         & 15.69                    & 1.14   & 0.001         & 0.50                           \\
Window-Averaged  (Ours)        & \textbf{15.10}       & \textbf{1.09}  &  \textbf{0.145}    & \textbf{0.56}                  \\
Instantaneous  (Ours)          & 15.50                   & 1.10  & 0.024  & 0.51                          \\
\bottomrule
\end{tabular}
\label{table:hmi_decisions_performance_test}
\end{table}

We next proceed to probe the efficacy of the resulting latent space to inform HMI adaptation to the subjects. We again used leave-one-out in the same manner to estimate the model's inference on the population, for both latent factor estimation and the resulting decision classifier based on the latent factors. In order to evaluate the interface selection decisions by the decision classifier we compared them to fixed interface choice chosen optimally for all participants (one-size-fits-all approach). We then measured the participants' behavior in terms of our chosen behavior statistic (mean average speed when yellow light was active) for the selected HMI choice (the classifier's decision) for the withheld subject averaged over the trials in which the experimental condition matched the decision classifiers output (thereby treating the experiment as a within-subject randomized trial study). 

\begin{figure}
    \centering
    \includegraphics[width=.9\textwidth, trim=50 120 50 130, clip]{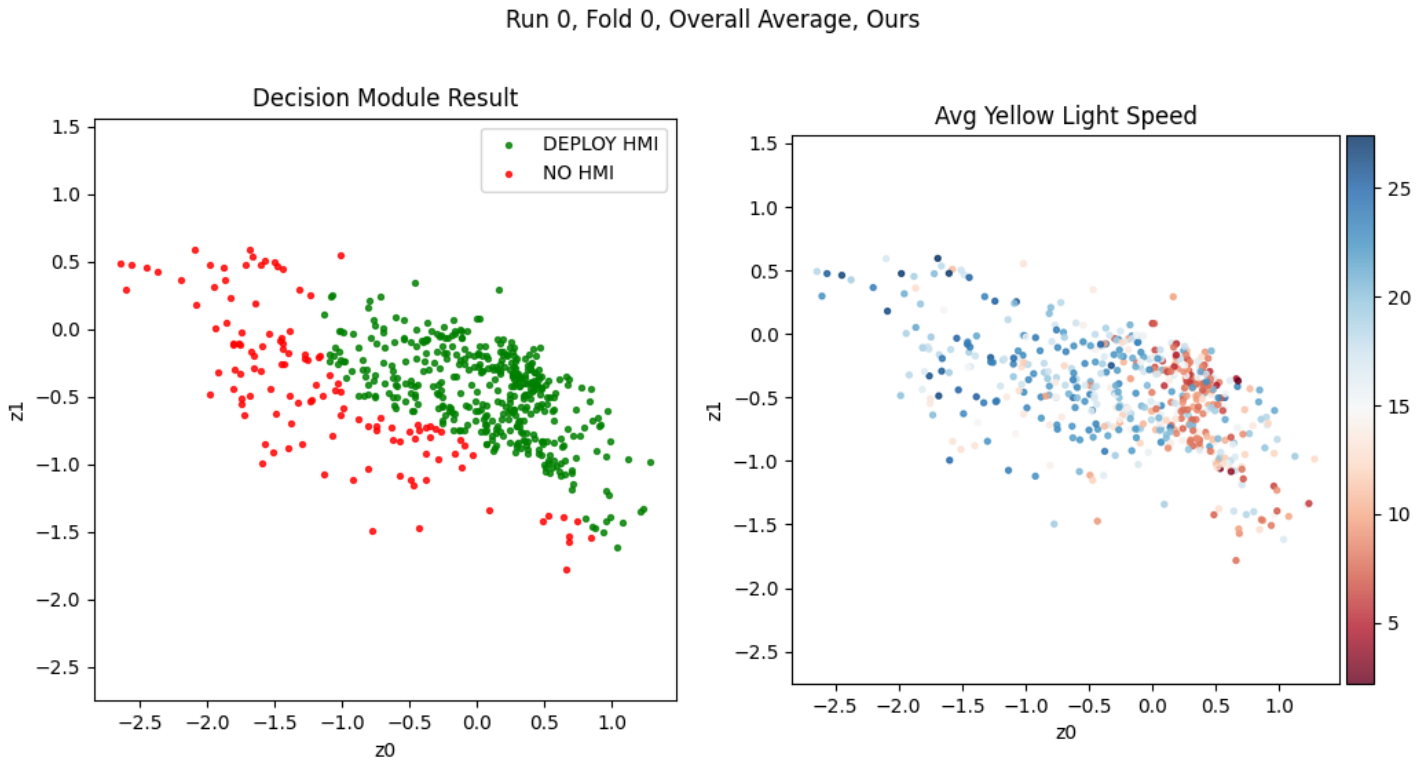}
    \caption{Trained decision boundary (left) and average speed during the yellow light phase conditioned on the decision scheme (right), plotted on the latent embedding space $z_0$, $z_1$.  Each point represents a unique time window over which the inference was run.}
    \label{fig:latent_factorsb}
\end{figure}

When leveraging the latent factors to decide on an HMI choice, we achieve a balanced accuracy of of 56\% and a Cohen Kappa of 0.145 in selecting the optimal HMI for the specific driver, as shown in Table~\ref{table:hmi_decisions_performance_test}, resulting in a reduction of 0.59 m/s in the mean speed over the duration of the yellow-phase of the traffic light.  Additionally, in Fig.~\ref{fig:latent_factorsb} (left), we code each of the latents generated from the trajectory snippets according to the decision module's predictions. In conjunction with Fig.~\ref{fig:latent_factorsb} (right), the trajectory snippets for which deployment of the HMI was the decision, we see that the average speed \textit{after} the yellow light transitions is lower, showing the effectiveness of the HMI decision scheme.

\subsection*{Discussion}

As traffic accidents and violations frequently occur due to poor impulsivity and inhibitory control, it is important to create driver safety systems that can overcome these cognitive limitations on a personalized level. In this work, we present an approach to infer the individual's latent factor, the use it to decide when it is or is not appropriate to show a driver safety interface depending on someone's inferred impulsivity and inhibitory control. 

To create this approach, we conducted a driving study using a high-fidelity motion simulator to understand how cognitive factors affect people's responses to driver safety interfaces. Our study revealed that the prototype interfaces had differing effects on drivers based on their impulsivity levels. In particular, we observed that drivers with lower levels of impulsivity tended to slow down when exposed to the interfaces, while drivers with higher levels of impulsivity exhibited the opposite response. Indeed, previous research has shown that impulsive drivers are more likely to run yellow lights~\cite{Chein2011-ll}, although yellow lights were designed to warn drivers that they may need to slow down. Our study is the first to show that vehicle safety interfaces may also lead to unintended driving behavior responses for some drivers based on their impulsivity. 

Leveraging the data collected in the study, we trained a recurrent neural network that can infer cognitive traits and, based on these, decide whether or not to employ a driver safety interface. The results show that our decision-making scheme can infer latent factors that are compact, correlate with cognitive measures associated with impulsivity, and can be used effectively to select driver interfaces to improve driver behavior, resulting in lower speed at the zone of dilemma of yellow lights. Although previous work has shown the relationship between cognitive factors such as impulsivity and driving behavior, this is the first time a model is proposed and examined so as to make driver safety recommendations based on cognitive factor inferences conditioned on the driver's behavior. 
 
The suggested approach lends itself to fleet-scale, online, in-vehicle optimization of the interaction with the driver across the population.  If deployed in such a manner, overall improvements in driver safety interfaces and safer roads could result.

\section*{Author contributions}
ES, JD, JC, DG, EK, SH, AM, AB, DB, HY, KS, TC, AB, and GR designed the research. ES, JD, JC, EG, EK, AM, AB, HN, DB, and HY performed the research. JD, DG, HN, BH , AP, and DB designed analytic tools. JD, DG, JC, and EK analyzed the data. ES, JD, JC, DG, EK, AM, HY, TC and GR wrote the paper. All authors reviewed the manuscript

\section*{Additional information}
All authors work for and receive compensation from Toyota Research Institute.

\section*{Data availability} Data and material will be made available upon request.


\end{document}


\flushbottom
\section{Supplemental methods}
\subsection*{Recruitment} 
39 external participants were locally recruited through a research partner. Participants were recruited based on their: (1) age, (2) gender, (3) driving experience, and (4) COVID-19 vaccination status. Equal participation based on age groups of interest and gender were used to counterbalance the distribution. Age, ethnicity and race were also taken into consideration during the recruitment process for a more robust and representative sample. Exclusionary criteria in the recruitment process included: (1) inactive driver’s license, (2) being pregnant. Participants of interest were contacted by the recruiter for an initial telephone screening. Approved participants would be contacted again by the recruiter and offered an incentive of  \$150 for the two-hour session. Participants who agreed to participate would then be contacted by email to establish a date and time for their session. Participants would be sent a confirmation email before their session date which would also contain a consent form with explicit instructions of what was to be expected of them.

\subsubsection*{Protocol}
Three participants were scheduled per testing day. Each session would last approximately two hours. Participants would be greeted by experimental proctors upon arrival and asked to complete an non-disclosure agreement as well as a COVID test before proceeding. Once a negative test result was acquired, along with a completed agreement, the subject would be led into the main testing room which includes the high fidelity driving simulator and operations station. Upon entering this space, participants would be seated on a couch in a sectioned off intake area. The proctor would then provide the subject with the same consent form sent prior to them through email but in a physical format. The subject would be given time to read over the form independently in the sectioned intake area while the proctors prepared the software for the simulated driving task. If the subject had questions, proctors would be nearby to clarify. When consent was granted by the subject, the proctor would collect and inspect the consent form before briefing the next part of the procedure. A proctor would store the consent form while another proctor would set up a laptop in the intake area loaded with a Qualtrics pre-driving survey to be completed. This survey would contain a few demographic questions and short questionnaires (see measures). After completing the survey, the proctor would collect the testing laptop and bring a charged E4 empatica wristband.

\section{Supplemental Results}
\subsection*{Effect of the human-machine interfaces on driving behavior}
\begin{figure}
    \centering
        \includegraphics[width=\linewidth]{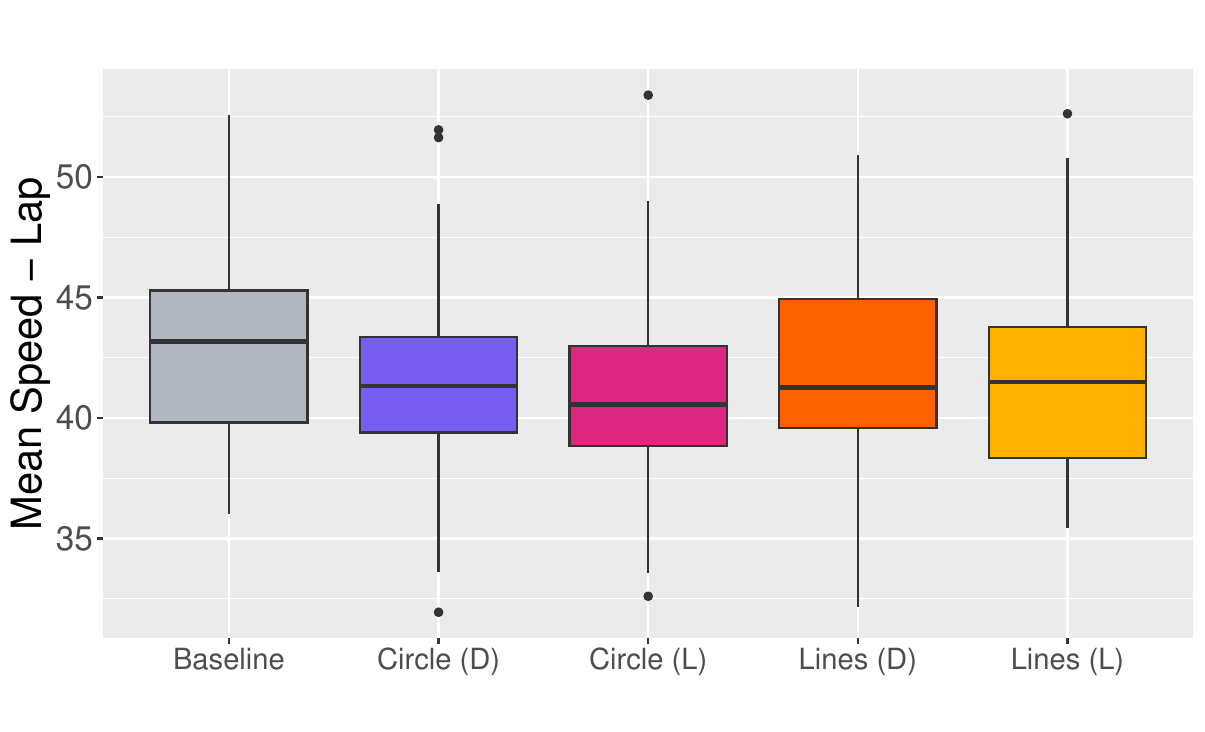}
    \caption{Mean speed during the lap in each condition. In all laps with HMI, participants had a lower mean speed during the lap in comparison to the baseline lap}
\label{fig:mean_speed_lap}
\end{figure}

We fitted separate linear mixed models to predict each driving behavior measure based on interface condition. The effect of all conditions is statistically significant and negative on the mean speed during the lap, as shown in Figure~\ref{fig:mean_speed_lap}.

\subsection*{Model Details}
In this section we describe in detail how we train and evaluate the context encoder and the decision classifier model.
\subsubsection*{Training Dataset}
In our work $\tau$ represents a trajectory of fixed number of samples $N$ of the states, where time is uniformly sampled from $0, \ldots, N$ consisting of observations of features and actions made by the driver (e.g. steering, throttling, braking). 
Specifically, the state ($s$) consists of a 5-dimensional vector comprised of the following features: longitudinal speed, distance to upcoming intersection start line, distance to upcoming intersection exit line, the state of the upcoming traffic light and whether the HMI was active or not. The action ($a$) consists of the longitudinal acceleration. 

The dataset we use for context encoder training consists of trajectory snippets ($\sim$22k) extracted around green to yellow light traffic light transitions (with window size equal to the context length used by the LSTM and a hopsize of 1) from all (baseline and HMI) laps from all the subjects. 

For the context encoder we use a history of driving behavior (temporally extended state and actions pairs) as inputs. Hyperparameter details are presented in Table~\ref{tab:hyperparameters}. The encoder structure consists of a single hidden layer LSTM with 128 units that maps the state and action history to a 2-dimensional latent space. The LSTM uses a context length of 30 timesteps which amounts to 6 seconds of driving behavior at 5 Hz. The hidden state of the LSTM is mapped to the mean and log of the standard deviation of the latent space. During each network training update, we conduct a forward pass using a batch (batch size = 2048) of training samples of past driving history and cognitive measures followed by network parameter update via backpropagation using the Adam optimizer. 
We select loss coefficients, batch size, and training epochs to empirically achieve reasonable convergence in the overall loss without overfitting.


\begin{table}[h]
\centering
\caption{Hyperparameters used in the training experiments.\label{tab:hyperparameters}}
\begin{tabular}{lcclc}
\toprule
Hyperparameter                                    & Value & \hspace{2cm}   & Hyperparameter                                    & Value   \\
\midrule
Batch size                                        & 2048    && Latent dimension                                  & 2 \\
Number of training epochs                         & 600     && Reconstruction loss coefficient $\alpha_1$           & $10^4$ \\
Contrastive margin $\epsilon$                     & 2       && Contrastive loss coefficient $\alpha_2$           & $10^4$ \\
Learning rate                                     & $10^{-2}$ && Latent regularizer loss coefficient $\alpha_3$    & $10^{-8}$ \\
LSTM context length                               & 6 s     && Support vector regression polynomial degree       & 3\\
LSTM hidden layers                                & 1       && Support vector regression margin $\epsilon_{SVR}$ & $0.5$ \\
LSTM hidden size                                  & 128 && &     \\ 
\bottomrule
\end{tabular}

\end{table}

\subsubsection*{Decision Classifier}
Our decision model uses an $\epsilon$-Support Vector Regression model with a polynomial kernel of degree 3 to regress over the continuous-valued training targets. The continuous-valued training target for each sample $z$ corresponds to the difference in the mean speeds when yellow lights are active between laps with and without HMI for that subject. 
Classification is done as a seprate step, with the decision threshold set to zero. The input to the decision model is the two-dimensional latent vector, $z$ generated by the trained context encoder on a dataset that only consists of trajectory snippets just preceding a green-to-yellow transition.

For the main reported results, we train the decision model only on latent vectors generated using the pre-trained context encoder from trajectory snippets that lead up to green to yellow traffic lights transitions sampled from all the laps. We filtered out snippets in which the subject was only exposed to an upcoming yellow light for less than 1s. The final dataset consists of 555 snippets extracted from 135 laps driven by 27 subjects. Out of the 555 snippets, 464 are from laps in which HMIs were deployed and the remaining 91 were from non-HMI laps.

\subsubsection*{Latent Factor Inference Modes}
The results in Table 3 included two schemes chosen for processing $z$'s as inputs to the decision classifier, both in training and inference, which we now expand upon.  For the \textit{Instantaneous} case shown in Table 3, the entire trajectory snippet is passed through the LSTM encoder and we consider $z$ to be obtained from the hidden state at only the last time step of the snippet.  We also explored a \textit{Windowed-Average} case, which we hypothesize better captures globally stable properties (for example, cognitive measures) of a particular driver.  To do this, we perform temporal averaging of the $z$'s generated at every timestep in the snippet over the entirety of the snippet.  
For each training configuration, we train the model 10 times using different random seeds and report the averaged results.

\subsection*{Effect of Sequential Inference on Model Performance}
To evaluate the performance of the context encoder and decision classifier as it might be deployed in a real world setting with streaming data, we execute the decision maker using all data fed sequentially into the model. Note that the LSTM context encoder is a recurrent model, so it will aggregate an estimate of $z$ in a way that is dependent on the history of driving behavior.  In Table~\ref{tab:hmi_decisions_performance_test_sequential_with_action_pred}, we observe similar trends to Table 3 and that our system outperforms the baselines, however, the improvement is milder.  In particular, Table 3 showed 0.59 m/s (standard error=1.58) reduction in yellow-light speed using Windowed-Average scheme, though when using streaming data, this reduction drops to 0.23 m/s (standard error=1.52) when comparing the same decision rules.  It is hypothesized that this drop in speed reduction may have occurred due to the noise introduced when using streaming data, and the fact that the LSTM retained state from data outside of the training context (e.g. periods of the drive when the lights were not visible to the driver).  Nonetheless, we can see that the inferred latent factors enable personalized HMI selection with 55\% accuracy, resulting in a mean yellow light speed of 15.23 m/s and decrease of 0.24 m/s compared to the globally-optimal Always-HMI case.

\begin{table}[]
\centering
\caption{Resulting accuracy of interface selection based on sequentially predicted latent factors on streaming data, averaged over 5 random seeds for the testing set over 27 folds. We use a moving average filter (of window length of 6s) to smooth out the sequential latent factor estimate time series. Last two rows of show results from an ablation in which the reconstruction loss was set to be 0.0}
\begin{tabular}{lcccc}
\toprule
\multirow{2}{*}
{Decision Rule} & \multicolumn{2}{c}{\begin{tabular}[c]{@{}c@{}}Mean Yellow-Light \\ Speed (m/s)\end{tabular}} & \multirow{2}{*}{\begin{tabular}[c]{@{}c@{}}Cohen's Kappa \\ Score\end{tabular}} & \multirow{2}{*}{\begin{tabular}[c]{@{}c@{}}Balanced \\ Accuracy\end{tabular}} \\
                               & $\mu$                    & Standard Error               &                           &                                    \\
\midrule
No-HMI                         & 17.36               & 1.12               & 0.0                       & 0.50                                \\
Always-HMI                     & 15.48                & 1.10               & 0.0            & 0.50                                \\
Random                         & 15.47                & 1.08              & -0.036                      & 0.47                           \\
Window-Averaged (Ours) - $\alpha_1 = 10^4$         & 15.28       & 1.09                 & 0.086                   & 0.54                  \\
Instantaneous (Ours) - $\alpha_1 = 10^4$           & 15.46                & 1.10       & 0.006           & 0.50                          \\
Window-Averaged (Ours) - $\alpha_1 = 0.0$       & \textbf{15.24}       & \textbf{1.07}                 & \textbf{0.108}                  & \textbf{0.55}                  \\
Instantaneous (Ours) - $\alpha_1 = 0.0$       & 15.46                & 1.09       & 0.071           & 0.52                          \\
\bottomrule
\end{tabular}
\label{tab:hmi_decisions_performance_test_sequential_with_action_pred}
\end{table}

\begin{table}[ht]
	\centering
	\caption{Go/No-Go Measures - Bayesian Pearson Correlations}
	\label{tab:bayesianPearsonCorrelations}
	{
		\begin{tabular}{lrrrrrr}
			\toprule
			 &  &  & Pearson's r & BF$_{1}$$_{0}$ & Lower 95\% CI & Upper 95\% CI  \\
			\cmidrule[0.4pt]{1-7}
			gonogo\_comission\_errors & - & gonogo\_comission\_errors\_rate & $0.690***$ & $5.060\times10^{+16}$ & $0.582$ & $0.767$  \\
			 & - & gonogo\_omission\_errors & $0.475***$ & $1.078\times10^{+6}$ & $0.325$ & $0.594$  \\
			 & - & gonogo\_average\_rt & $0.041$ & $0.122$ & $-0.131$ & $0.210$  \\
			 & - & mean\_speed\_yellow & $0.107$ & $0.225$ & $-0.066$ & $0.271$  \\
			 & - & std\_speed\_yellow & $-0.062$ & $0.139$ & $-0.229$ & $0.111$  \\
			 & - & max\_speed\_yellow & $0.135$ & $0.352$ & $-0.037$ & $0.297$  \\
			 & - & min\_speed\_yellow & $0.071$ & $0.151$ & $-0.101$ & $0.238$  \\
			gonogo\_comission\_errors\_rate & - & gonogo\_omission\_errors & $-0.131$ & $0.326$ & $-0.293$ & $0.042$  \\
			 & - & gonogo\_average\_rt & $-0.414***$ & $15714.643$ & $-0.543$ & $-0.257$  \\
			 & - & mean\_speed\_yellow & $0.348***$ & $375.156$ & $0.184$ & $0.487$  \\
			 & - & std\_speed\_yellow & $-0.134$ & $0.347$ & $-0.297$ & $0.038$  \\
			 & - & max\_speed\_yellow & $0.360***$ & $681.940$ & $0.197$ & $0.497$  \\
			 & - & min\_speed\_yellow & $0.172$ & $0.727$ & $-4.975\times10^{-4}$ & $0.331$  \\
			gonogo\_omission\_errors & - & gonogo\_average\_rt & $0.525***$ & $7.110\times10^{+7}$ & $0.383$ & $0.635$  \\
			 & - & mean\_speed\_yellow & $-0.093$ & $0.189$ & $-0.258$ & $0.080$  \\
			 & - & std\_speed\_yellow & $0.002$ & $0.110$ & $-0.169$ & $0.172$  \\
			 & - & max\_speed\_yellow & $-0.148$ & $0.446$ & $-0.309$ & $0.024$  \\
			 & - & min\_speed\_yellow & $-0.032$ & $0.117$ & $-0.201$ & $0.140$  \\
			gonogo\_average\_rt & - & mean\_speed\_yellow & $-0.460***$ & $352747.434$ & $-0.581$ & $-0.309$  \\
			 & - & std\_speed\_yellow & $0.245$ & $5.482$ & $0.075$ & $0.397$  \\
			 & - & max\_speed\_yellow & $-0.406***$ & $9205.034$ & $-0.536$ & $-0.248$  \\
			 & - & min\_speed\_yellow & $-0.310**$ & $63.828$ & $-0.454$ & $-0.143$  \\
			mean\_speed\_yellow & - & std\_speed\_yellow & $-0.684***$ & $9.245\times10^{+16}$ & $-0.761$ & $-0.577$  \\
			 & - & max\_speed\_yellow & $0.706***$ & $4.952\times10^{+18}$ & $0.605$ & $0.779$  \\
			 & - & min\_speed\_yellow & $0.809***$ & $1.845\times10^{+29}$ & $0.737$ & $0.858$  \\
			std\_speed\_yellow & - & max\_speed\_yellow & $-0.103$ & $0.218$ & $-0.265$ & $0.066$  \\
			 & - & min\_speed\_yellow & $-0.911***$ & $2.642\times10^{+49}$ & $-0.935$ & $-0.874$  \\
			max\_speed\_yellow & - & min\_speed\_yellow & $0.362***$ & $1081.461$ & $0.203$ & $0.496$  \\
			\bottomrule
			\multicolumn{7}{p{0.5\linewidth}}{*  BF$_{1}$$_{0}$  > 10, ** BF$_{1}$$_{0}$  > 30, *** BF$_{1}$$_{0}$  > 100} \\
		\end{tabular}
	}
\end{table}

\begin{table}[h]
	\centering
	\caption{Stop Signal Measures - Bayesian  Pearson Correlations}
	\label{tab:bayesianStopSignalCorrelations}
	{
		\begin{tabular}{lrrrrrr}
			\toprule
			 &  &  & Pearson's r & BF$_{1}$$_{0}$ & Lower 95\% CI & Upper 95\% CI  \\
			\cmidrule[0.4pt]{1-7}
			ssd & - & ssrt & $-0.366***$ & $1347.018$ & $-0.500$ & $-0.207$  \\
			 & - & usRT & $0.857***$ & $6.229\times10^{+36}$ & $0.801$ & $0.895$  \\
			 & - & goRT\_all & $0.954***$ & $3.339\times10^{+67}$ & $0.934$ & $0.966$  \\
			 & - & goRT\_sd & $0.730***$ & $5.189\times10^{+20}$ & $0.635$ & $0.798$  \\
			 & - & go\_omission & $0.736***$ & $1.628\times10^{+21}$ & $0.642$ & $0.802$  \\
			 & - & mean\_speed\_yellow & $-0.349***$ & $524.864$ & $-0.485$ & $-0.188$  \\
			 & - & std\_speed\_yellow & $0.269*$ & $15.043$ & $0.104$ & $0.415$  \\
			 & - & max\_speed\_yellow & $-0.155$ & $0.533$ & $-0.313$ & $0.014$  \\
			 & - & min\_speed\_yellow & $-0.274*$ & $18.012$ & $-0.420$ & $-0.109$  \\
			ssrt & - & usRT & $0.036$ & $0.117$ & $-0.132$ & $0.202$  \\
			 & - & goRT\_all & $-0.138$ & $0.381$ & $-0.297$ & $0.031$  \\
			 & - & goRT\_sd & $-0.342***$ & $366.723$ & $-0.479$ & $-0.181$  \\
			 & - & go\_omission & $-0.131$ & $0.333$ & $-0.290$ & $0.039$  \\
			 & - & mean\_speed\_yellow & $0.042$ & $0.121$ & $-0.127$ & $0.207$  \\
			 & - & std\_speed\_yellow & $-0.021$ & $0.111$ & $-0.188$ & $0.147$  \\
			 & - & max\_speed\_yellow & $-0.050$ & $0.127$ & $-0.215$ & $0.119$  \\
			 & - & min\_speed\_yellow & $-1.106\times10^{-4}$ & $0.108$ & $-0.167$ & $0.167$  \\
			usRT & - & goRT\_all & $0.965***$ & $3.003\times10^{+75}$ & $0.950$ & $0.975$  \\
			 & - & goRT\_sd & $0.560***$ & $5.786\times10^{+9}$ & $0.428$ & $0.662$  \\
			 & - & go\_omission & $0.665***$ & $4.605\times10^{+15}$ & $0.554$ & $0.747$  \\
			 & - & mean\_speed\_yellow & $-0.355***$ & $731.749$ & $-0.490$ & $-0.195$  \\
			 & - & std\_speed\_yellow & $0.209$ & $1.999$ & $0.041$ & $0.361$  \\
			 & - & max\_speed\_yellow & $-0.246$ & $6.393$ & $-0.394$ & $-0.079$  \\
			 & - & min\_speed\_yellow & $-0.232$ & $4.076$ & $-0.382$ & $-0.065$  \\
			goRT\_all & - & goRT\_sd & $0.686***$ & $1.251\times10^{+17}$ & $0.579$ & $0.763$  \\
			 & - & go\_omission & $0.720***$ & $7.035\times10^{+19}$ & $0.623$ & $0.790$  \\
			 & - & mean\_speed\_yellow & $-0.380***$ & $2933.574$ & $-0.512$ & $-0.222$  \\
			 & - & std\_speed\_yellow & $0.260$ & $10.406$ & $0.093$ & $0.407$  \\
			 & - & max\_speed\_yellow & $-0.226$ & $3.384$ & $-0.377$ & $-0.059$  \\
			 & - & min\_speed\_yellow & $-0.277$ & $19.944$ & $-0.422$ & $-0.111$  \\
			goRT\_sd & - & go\_omission & $0.576***$ & $3.075\times10^{+10}$ & $0.446$ & $0.675$  \\
			 & - & mean\_speed\_yellow & $-0.032$ & $0.115$ & $-0.198$ & $0.136$  \\
			 & - & std\_speed\_yellow & $-0.012$ & $0.109$ & $-0.178$ & $0.156$  \\
			 & - & max\_speed\_yellow & $0.033$ & $0.116$ & $-0.135$ & $0.199$  \\
			 & - & min\_speed\_yellow & $0.042$ & $0.121$ & $-0.127$ & $0.207$  \\
			go\_omission & - & mean\_speed\_yellow & $-0.233$ & $4.179$ & $-0.383$ & $-0.066$  \\
			 & - & std\_speed\_yellow & $0.097$ & $0.199$ & $-0.073$ & $0.259$  \\
			 & - & max\_speed\_yellow & $-0.164$ & $0.645$ & $-0.321$ & $0.005$  \\
			 & - & min\_speed\_yellow & $-0.070$ & $0.148$ & $-0.233$ & $0.100$  \\
			mean\_speed\_yellow & - & std\_speed\_yellow & $-0.684***$ & $9.245\times10^{+16}$ & $-0.761$ & $-0.577$  \\
			 & - & max\_speed\_yellow & $0.706***$ & $4.952\times10^{+18}$ & $0.605$ & $0.779$  \\
			 & - & min\_speed\_yellow & $0.809***$ & $1.845\times10^{+29}$ & $0.737$ & $0.858$  \\
			std\_speed\_yellow & - & max\_speed\_yellow & $-0.103$ & $0.218$ & $-0.265$ & $0.066$  \\
			 & - & min\_speed\_yellow & $-0.911***$ & $2.642\times10^{+49}$ & $-0.935$ & $-0.874$  \\
			max\_speed\_yellow & - & min\_speed\_yellow & $0.362***$ & $1081.461$ & $0.203$ & $0.496$  \\
			\bottomrule
			\multicolumn{7}{p{0.5\linewidth}}{*  BF$_{1}$$_{0}$  > 10, ** BF$_{1}$$_{0}$  > 30, *** BF$_{1}$$_{0}$  > 100} \\
		\end{tabular}
  
	}
\end{table}

\begin{table}[h]
	\centering
	\caption{BIS/BAS Measures - Bayesian Pearson Correlations}
	\label{tab:bayesianBISBASCorrelations}
	{
		\begin{tabular}{lrrrrrr}
			\toprule
			 &  &  & Pearson's r & BF$_{1}$$_{0}$ & Lower 95\% CI & Upper 95\% CI  \\
			\cmidrule[0.4pt]{1-7}
			BAS\_Drive & - & BAS\_Fun & $0.275*$ & $18.825$ & $0.110$ & $0.421$  \\
			 & - & BAS\_reward & $-0.065$ & $0.142$ & $-0.229$ & $0.104$  \\
			 & - & BIS & $-3.527\times10^{-4}$ & $0.108$ & $-0.168$ & $0.167$  \\
			 & - & mean\_speed\_yellow & $0.168$ & $0.709$ & $-7.937\times10^{-4}$ & $0.325$  \\
			 & - & std\_speed\_yellow & $-0.242$ & $5.589$ & $-0.391$ & $-0.075$  \\
			 & - & max\_speed\_yellow & $-0.043$ & $0.121$ & $-0.208$ & $0.126$  \\
			 & - & min\_speed\_yellow & $0.220$ & $2.770$ & $0.052$ & $0.371$  \\
			BAS\_Fun & - & BAS\_reward & $0.554***$ & $2.848\times10^{+9}$ & $0.420$ & $0.657$  \\
			 & - & BIS & $-0.206$ & $1.823$ & $-0.358$ & $-0.037$  \\
			 & - & mean\_speed\_yellow & $0.473***$ & $1.700\times10^{+6}$ & $0.326$ & $0.590$  \\
			 & - & std\_speed\_yellow & $-0.329***$ & $196.005$ & $-0.468$ & $-0.167$  \\
			 & - & max\_speed\_yellow & $0.315**$ & $99.190$ & $0.152$ & $0.455$  \\
			 & - & min\_speed\_yellow & $0.362***$ & $1075.032$ & $0.203$ & $0.496$  \\
			BAS\_reward & - & BIS & $0.242$ & $5.668$ & $0.075$ & $0.391$  \\
			 & - & mean\_speed\_yellow & $0.167$ & $0.681$ & $-0.003$ & $0.323$  \\
			 & - & std\_speed\_yellow & $0.053$ & $0.129$ & $-0.116$ & $0.217$  \\
			 & - & max\_speed\_yellow & $0.294**$ & $39.632$ & $0.129$ & $0.437$  \\
			 & - & min\_speed\_yellow & $0.038$ & $0.118$ & $-0.131$ & $0.203$  \\
			BIS & - & mean\_speed\_yellow & $-0.104$ & $0.221$ & $-0.266$ & $0.065$  \\
			 & - & std\_speed\_yellow & $0.102$ & $0.212$ & $-0.068$ & $0.263$  \\
			 & - & max\_speed\_yellow & $-0.044$ & $0.122$ & $-0.209$ & $0.125$  \\
			 & - & min\_speed\_yellow & $-0.110$ & $0.240$ & $-0.271$ & $0.059$  \\
			mean\_speed\_yellow & - & std\_speed\_yellow & $-0.684***$ & $9.245\times10^{+16}$ & $-0.761$ & $-0.577$  \\
			 & - & max\_speed\_yellow & $0.706***$ & $4.952\times10^{+18}$ & $0.605$ & $0.779$  \\
			 & - & min\_speed\_yellow & $0.809***$ & $1.845\times10^{+29}$ & $0.737$ & $0.858$  \\
			std\_speed\_yellow & - & max\_speed\_yellow & $-0.103$ & $0.218$ & $-0.265$ & $0.066$  \\
			 & - & min\_speed\_yellow & $-0.911***$ & $2.642\times10^{+49}$ & $-0.935$ & $-0.874$  \\
			max\_speed\_yellow & - & min\_speed\_yellow & $0.362***$ & $1081.461$ & $0.203$ & $0.496$  \\
			\bottomrule
			\multicolumn{7}{p{0.5\linewidth}}{*  BF$_{1}$$_{0}$  > 10, ** BF$_{1}$$_{0}$  > 30, *** BF$_{1}$$_{0}$  > 100} \\
		\end{tabular}
	}
\end{table}

\begin{table}[h]
	\centering
	\caption{UPPS-P Bayesian Pearson Correlations}
	\label{tab:bayesianUPPSPCorrelations}
	{
		\begin{tabular}{lrrrrrr}
			\toprule
			 &  &  & Pearson's r & BF$_{1}$$_{0}$ & Lower 95\% CI & Upper 95\% CI  \\
			\cmidrule[0.4pt]{1-7}
			uppsp\_negative\_urgency & - & uppsp\_lack\_perseverance & $0.257$ & $9.623$ & $0.091$ & $0.405$  \\
			 & - & uppsp\_lack\_premeditation & $0.396***$ & $7487.515$ & $0.240$ & $0.525$  \\
			 & - & uppsp\_sensation\_seeking & $-0.015$ & $0.109$ & $-0.181$ & $0.153$  \\
			 & - & uppsp\_positive\_urgency & $0.751***$ & $5.046\times10^{+22}$ & $0.662$ & $0.814$  \\
			 & - & mean\_speed\_yellow & $0.026$ & $0.113$ & $-0.142$ & $0.192$  \\
			 & - & std\_speed\_yellow & $0.029$ & $0.114$ & $-0.139$ & $0.195$  \\
			 & - & max\_speed\_yellow & $0.100$ & $0.207$ & $-0.070$ & $0.261$  \\
			 & - & min\_speed\_yellow & $-0.079$ & $0.163$ & $-0.243$ & $0.090$  \\
			uppsp\_lack\_perseverance & - & uppsp\_lack\_premeditation & $0.253$ & $8.075$ & $0.086$ & $0.400$  \\
			 & - & uppsp\_sensation\_seeking & $-0.200$ & $1.544$ & $-0.353$ & $-0.031$  \\
			 & - & uppsp\_positive\_urgency & $0.275*$ & $18.855$ & $0.110$ & $0.421$  \\
			 & - & mean\_speed\_yellow & $-0.010$ & $0.108$ & $-0.177$ & $0.157$  \\
			 & - & std\_speed\_yellow & $0.073$ & $0.152$ & $-0.097$ & $0.236$  \\
			 & - & max\_speed\_yellow & $0.108$ & $0.233$ & $-0.061$ & $0.269$  \\
			 & - & min\_speed\_yellow & $-0.105$ & $0.224$ & $-0.267$ & $0.064$  \\
			uppsp\_lack\_premeditation & - & uppsp\_sensation\_seeking & $0.284*$ & $26.998$ & $0.119$ & $0.429$  \\
			 & - & uppsp\_positive\_urgency & $0.279*$ & $21.447$ & $0.113$ & $0.424$  \\
			 & - & mean\_speed\_yellow & $0.189$ & $1.158$ & $0.020$ & $0.343$  \\
			 & - & std\_speed\_yellow & $0.012$ & $0.109$ & $-0.156$ & $0.178$  \\
			 & - & max\_speed\_yellow & $0.240$ & $5.282$ & $0.073$ & $0.389$  \\
			 & - & min\_speed\_yellow & $-0.021$ & $0.111$ & $-0.187$ & $0.147$  \\
			uppsp\_sensation\_seeking & - & uppsp\_positive\_urgency & $0.109$ & $0.236$ & $-0.060$ & $0.270$  \\
			 & - & mean\_speed\_yellow & $0.295**$ & $42.896$ & $0.131$ & $0.438$  \\
			 & - & std\_speed\_yellow & $0.110$ & $0.237$ & $-0.060$ & $0.271$  \\
			 & - & max\_speed\_yellow & $0.471***$ & $1.540\times10^{+6}$ & $0.325$ & $0.589$  \\
			 & - & min\_speed\_yellow & $0.024$ & $0.112$ & $-0.144$ & $0.190$  \\
			uppsp\_positive\_urgency & - & mean\_speed\_yellow & $0.059$ & $0.135$ & $-0.110$ & $0.223$  \\
			 & - & std\_speed\_yellow & $0.131$ & $0.337$ & $-0.038$ & $0.291$  \\
			 & - & max\_speed\_yellow & $0.284*$ & $26.932$ & $0.119$ & $0.429$  \\
			 & - & min\_speed\_yellow & $-0.116$ & $0.261$ & $-0.277$ & $0.054$  \\
			mean\_speed\_yellow & - & std\_speed\_yellow & $-0.684***$ & $9.245\times10^{+16}$ & $-0.761$ & $-0.577$  \\
			 & - & max\_speed\_yellow & $0.706***$ & $4.952\times10^{+18}$ & $0.605$ & $0.779$  \\
			 & - & min\_speed\_yellow & $0.809***$ & $1.845\times10^{+29}$ & $0.737$ & $0.858$  \\
			std\_speed\_yellow & - & max\_speed\_yellow & $-0.103$ & $0.218$ & $-0.265$ & $0.066$  \\
			 & - & min\_speed\_yellow & $-0.911***$ & $2.642\times10^{+49}$ & $-0.935$ & $-0.874$  \\
			max\_speed\_yellow & - & min\_speed\_yellow & $0.362***$ & $1081.461$ & $0.203$ & $0.496$  \\
			\bottomrule
			\addlinespace[1ex]
			\multicolumn{7}{p{0.5\linewidth}}{*  BF$_{1}$$_{0}$  > 10, ** BF$_{1}$$_{0}$  > 30, *** BF$_{1}$$_{0}$  > 100} \\
		\end{tabular}
	}
\end{table}

\begin{table}[h]
	\centering
	\caption{Manchester DBQ Measures - Bayesian Pearson Correlations}
	\label{tab:bayesianDBQCorrelations}
	{
		\begin{tabular}{lrrrrrr}
			\toprule
			 &  &  & Pearson's r & BF$_{1}$$_{0}$ & Lower 95\% CI & Upper 95\% CI  \\
			\cmidrule[0.4pt]{1-7}
			dbq\_aggressiveViolations & - & dbq\_errors & $0.468***$ & $1.139\times10^{+6}$ & $0.320$ & $0.586$  \\
			 & - & dbq\_ordinaryViolations & $0.490***$ & $7.218\times10^{+6}$ & $0.346$ & $0.605$  \\
			 & - & dbq\_lapses & $0.176$ & $0.844$ & $0.007$ & $0.332$  \\
			 & - & mean\_speed\_yellow & $0.040$ & $0.120$ & $-0.128$ & $0.206$  \\
			 & - & std\_speed\_yellow & $-0.021$ & $0.111$ & $-0.187$ & $0.147$  \\
			 & - & max\_speed\_yellow & $0.070$ & $0.149$ & $-0.099$ & $0.234$  \\
			 & - & min\_speed\_yellow & $0.018$ & $0.110$ & $-0.150$ & $0.185$  \\
			dbq\_errors & - & dbq\_ordinaryViolations & $0.577***$ & $3.647\times10^{+10}$ & $0.447$ & $0.676$  \\
			 & - & dbq\_lapses & $0.711***$ & $1.084\times10^{+19}$ & $0.610$ & $0.782$  \\
			 & - & mean\_speed\_yellow & $0.163$ & $0.624$ & $-0.007$ & $0.320$  \\
			 & - & std\_speed\_yellow & $0.019$ & $0.110$ & $-0.149$ & $0.185$  \\
			 & - & max\_speed\_yellow & $0.251$ & $7.678$ & $0.084$ & $0.399$  \\
			 & - & min\_speed\_yellow & $0.011$ & $0.109$ & $-0.156$ & $0.178$  \\
			dbq\_ordinaryViolations & - & dbq\_lapses & $0.337***$ & $286.333$ & $0.175$ & $0.474$  \\
			 & - & mean\_speed\_yellow & $0.400***$ & $9693.128$ & $0.244$ & $0.529$  \\
			 & - & std\_speed\_yellow & $-0.056$ & $0.133$ & $-0.221$ & $0.113$  \\
			 & - & max\_speed\_yellow & $0.545***$ & $1.141\times10^{+9}$ & $0.409$ & $0.650$  \\
			 & - & min\_speed\_yellow & $0.188$ & $1.134$ & $0.019$ & $0.342$  \\
			dbq\_lapses & - & mean\_speed\_yellow & $0.038$ & $0.119$ & $-0.130$ & $0.204$  \\
			 & - & std\_speed\_yellow & $-0.050$ & $0.127$ & $-0.215$ & $0.119$  \\
			 & - & max\_speed\_yellow & $0.011$ & $0.108$ & $-0.157$ & $0.177$  \\
			 & - & min\_speed\_yellow & $8.082\times10^{-4}$ & $0.108$ & $-0.166$ & $0.168$  \\
			mean\_speed\_yellow & - & std\_speed\_yellow & $-0.684***$ & $9.245\times10^{+16}$ & $-0.761$ & $-0.577$  \\
			 & - & max\_speed\_yellow & $0.706***$ & $4.952\times10^{+18}$ & $0.605$ & $0.779$  \\
			 & - & min\_speed\_yellow & $0.809***$ & $1.845\times10^{+29}$ & $0.737$ & $0.858$  \\
			std\_speed\_yellow & - & max\_speed\_yellow & $-0.103$ & $0.218$ & $-0.265$ & $0.066$  \\
			 & - & min\_speed\_yellow & $-0.911***$ & $2.642\times10^{+49}$ & $-0.935$ & $-0.874$  \\
			max\_speed\_yellow & - & min\_speed\_yellow & $0.362***$ & $1081.461$ & $0.203$ & $0.496$  \\
			\bottomrule
			\multicolumn{7}{p{0.5\linewidth}}{*  BF$_{1}$$_{0}$  > 10, ** BF$_{1}$$_{0}$  > 30, *** BF$_{1}$$_{0}$  > 100} \\
		\end{tabular}
	}
\end{table}

\begin{table}[h]
	\centering
	\caption{Demographics - Bayesian Pearson Correlations}
	\label{tab:bayesianDemographicsCorrelations}
	{
		\begin{tabular}{lrrrrrr}
			\toprule
			 &  &  & Pearson's r & BF$_{1}$$_{0}$ & Lower 95\% CI & Upper 95\% CI  \\
			\cmidrule[0.4pt]{1-7}
			exactAge & - & gender & $-0.487***$ & $5.519\times10^{+6}$ & $-0.602$ & $-0.343$  \\
			 & - & mean\_speed\_yellow & $-0.288**$ & $30.863$ & $-0.432$ & $-0.123$  \\
			 & - & std\_speed\_yellow & $0.074$ & $0.154$ & $-0.096$ & $0.237$  \\
			 & - & max\_speed\_yellow & $-0.269*$ & $14.804$ & $-0.415$ & $-0.103$  \\
			 & - & min\_speed\_yellow & $-0.104$ & $0.220$ & $-0.265$ & $0.066$  \\
			gender & - & mean\_speed\_yellow & $-0.013$ & $0.109$ & $-0.180$ & $0.154$  \\
			 & - & std\_speed\_yellow & $0.159$ & $0.582$ & $-0.010$ & $0.317$  \\
			 & - & max\_speed\_yellow & $0.050$ & $0.127$ & $-0.119$ & $0.215$  \\
			 & - & min\_speed\_yellow & $-0.151$ & $0.484$ & $-0.308$ & $0.019$  \\
			mean\_speed\_yellow & - & std\_speed\_yellow & $-0.684***$ & $9.245\times10^{+16}$ & $-0.761$ & $-0.577$  \\
			 & - & max\_speed\_yellow & $0.706***$ & $4.952\times10^{+18}$ & $0.605$ & $0.779$  \\
			 & - & min\_speed\_yellow & $0.809***$ & $1.845\times10^{+29}$ & $0.737$ & $0.858$  \\
			std\_speed\_yellow & - & max\_speed\_yellow & $-0.103$ & $0.218$ & $-0.265$ & $0.066$  \\
			 & - & min\_speed\_yellow & $-0.911***$ & $2.642\times10^{+49}$ & $-0.935$ & $-0.874$  \\
			max\_speed\_yellow & - & min\_speed\_yellow & $0.362***$ & $1081.461$ & $0.203$ & $0.496$  \\
			\bottomrule
			\multicolumn{7}{p{0.5\linewidth}}{*  BF$_{1}$$_{0}$  > 10, ** BF$_{1}$$_{0}$  > 30, *** BF$_{1}$$_{0}$  > 100} \\
		\end{tabular}
	}
\end{table}

\






